\documentclass[fleqn,usenatbib]{mnras}

\usepackage{newtxtext,newtxmath}

\usepackage[T1]{fontenc}

\usepackage{graphicx}	
\usepackage{amsmath}	
\usepackage{comment}

\newcommand{\nc}{\newcommand}
\newcommand{\HII}{H {\sc ii}}
\nc{\msun}{\ensuremath{\mathrm{M}_\odot}}
\nc{\lsun}{\ensuremath{\mathrm{L}_\odot}}
\nc{\thCO}{$^{13}$CO}
\nc{\CeiO}{C$^{18}$O}
\nc{\kms}{\mbox{km~s$^{-1}$}}
\nc{\Kkms}{\mbox{K\,km~s$^{-1}$}}

\nc{\twCO}{$^{12}$CO}
\newcommand\arcdeg{\mbox{$^\circ$}}%
\nc{\cmsq}{\mbox{cm$^{-2}$}}
\nc{\cmcub}{\mbox{cm$^{-3}$}}

\title[Star formation in G326.27-0.49]{Spiral Structure and Massive Star formation in the Hub-Filament-System G326.27-0.49}

\author[B. Mookerjea et al.]{
Bhaswati Mookerjea,$^{1}$\thanks{E-mail: bhaswati@tifr.res.in}
V. S. Veena,$^{2}$ Rolf G\"usten$^{2}$, F. Wyrowski$^{2}$, Akhil Lasrado$^{3}$
\\
$^{1}$Department of Astronomy \& Astrophysics, Tata Institute of
Fundamental Research, Homi Bhabha Road, Mumbai 400005, India\\
$^{2}$ Max Planck Institut f\"ur Radioastronomie, Auf dem H\"ugel 69, D-53121 Bonn, Germany \\
$^{3}$ Indian Institute of Science Education and Research, Kolkata, India\\
}

\date{Accepted 16 January 2023. Received 13 January 2023; in original form 10 October 2022}

\pubyear{2023}

\begin{document}
\label{firstpage}
\pagerange{\pageref{firstpage}--\pageref{lastpage}}
\maketitle

\begin{abstract}
Hub-filament systems (HFSs) are potential sites of formation of star
clusters and high mass stars. To understand the HFSs and to provide
observational constraints on current theories that attempt to explain
star formation globally, we report a study of the region associated with
G326.27-0.49 using infrared data of dust continuum and newly obtained
observations on molecular tracers using the APEX telescope. We use the
spectroscopic observations to identify velocity-coherent structures
(filaments and clumps) and study their properties at a resolution of
0.4\,pc. The region contains two main velocity components: first
component shows four filaments between -63 and -55\,\kms\ forming a
spiral structure converging in a hub, the second filamentary component
at $\sim -72$\,\kms\ harbors a massive young stellar object and possibly
interacts with the hub. The clumps harbouring the three main YSOs
in the region are massive (187--535\,\msun), have luminosities
consistent with B-type stars, have central densities of $\sim
10^6$\,\cmcub\ and drive large outflows.  Majority of the
velocity-coherent clumps in the region show virial parameters between
2--7, which considering the detection of protostars implies collapse to
be gradual.  We conclude that the region consists of a network of
filaments through which mass accretes ($\sim 10^{-4}$\,\msun\,
yr$^{-1}$) onto the hub.  The hub and some of the ends of filaments
appears to be undergoing collapse to form new stars. This study
identifies a target region for future high resolution studies that would
probe the link between the core and filament evolution.
\end{abstract}




\section{Introduction} 

Massive star formation remains a poorly understood phenomenon, largely
due to the difficulty of identifying and studying massive young stellar
objects (MYSOs) in the crucial early active accretion and outflow
phase. During the earliest stages of their evolution, young MYSOs
remain deeply embedded in their natal clouds. Most massive star-forming
regions are also distant ($>$1 kpc) and crowded, with massive stars
forming in close proximity to other MYSOs and a large numbers of lower
mass young stellar objects (YSOs). Studying the early stages of massive
star formation thus requires high angular resolution observations (to
resolve individual objects in crowded regions) at long wavelengths
unaffected by extinction.
Availability of improved observational facilities at long wavelength (infrared and beyond) such as Spitzer and Herschel have led to observations of a large number of molecular clouds
that have revealed a ubiquity of filamentary structures containing stars
in different evolutionary stages \citep[e.g.,][and others]{Schneider1979,Myers2009,Andre2010,Molinari2010,peretto2014}.  Filamentary structures pervading clouds are unstable
against both radial collapse and fragmentation \citep[e.g.,][]{Larson1985,Inutsuka1997}, and although their origin or formation process
is still unclear, turbulence and gravity \citep[e.g.,][]{Klessen2000,Andre2010} can
produce, together with the presence of magnetic fields
\citep[e.g.,][]{Molina2012,Kirk2015}, the observed structures. It is thought that star
formation occurs preferentially along the filaments, with high-mass stars forming in the
highest density regions where several filaments converge, called ridges or hubs
\citep[$N_{\rm H}\sim 10^{23}$\,\cmsq\ and $n_{\rm H_2}\sim 10^6$\,\cmcub, e.g.,][]{Schneider2010,Schneider2012,Peretto2013,peretto2014}. Additionally,
\citet{Myers2009} have presented evidence that multiple parsec-scale filaments tend to
branch out from "hubs" in regions forming stellar groups in both nearby and distant
regions.  This scenario also constitutes the backbone of some of the competing theories of
high mass star formation, e.g., the global hierarchical collapse (GHC) model, in which all size scales
are contracting gravitationally, and accreting from the next larger scale
\citep{VSemadeni2019}, the inertial-inflow model, in which most of the final stellar mass is channeled toward the accreting star by the random velocity field from large
scales, unaffected by the stellar gravity \citep{Padoan2020}. In the recent years several theoretical and observational studies have addressed the dynamics and fragmentation of filamentary structures \citep[e.g.,][]{Andre2010,Arzoumanian2019,Clarke2019}. However, few of these works focus on massive star-forming regions within hub-filament system (HFS), and little is known about the dynamics of filamentary networks (e.g., cluster-forming hub filament systems) and their role in the accretion processes that regulate the formation of high-mass star-forming clusters. Identification of such hub-filament-systems (HFS) in the large scale survey observations followed by a methodical analysis of the dynamics of the molecular material and the magnetic fields in such regions is thus a robust approach toward the study of the formation of MYSOs \citep[e.g.,][and references therein]{Trevino2019,Wang2020,Hacar2022}. 

The Hub-Filament System G326.27-0.49, located at a distance of 3.6\,kpc
\citep{elia2017} is associated with three filaments \citep{kumar2020}
and harbors two extended green objects (EGO G326.27-0.49 and EGO
G326.32-0.39) that are indicative of outflow activity from  embedded
protostellar objects \citep[EGO][]{cyganowski2008}. The region has been
observed as part of several continuum and spectroscopic survey
observations, but no detailed analysis of the molecular emission from
the region is available in the literature. So far no other signatures of
massive star formation such as \HII\ region or masers have been detected
in this region.  \citet{peretto2009} identified three Spitzer dark
clouds in close proximity of the EGO 326.27-0.49.  At 870\,\micron\ the ATLASGAL survey has detected four clumps in this region, three of which are associated with Hi-GAL sources visible at $\lambda \geq $70\,\micron. Additionally there are two Hi-GAL sources detected only at $\lambda >70$\,\micron\ (70\,\micron -dark). The continuum data reveal a population of protostellar objects at different stages of evolution in the region and in particular the presence of multiple sub-millimeter sources together with the hub seen at the confluence of multiple filaments makes the region interesting as a case study for the initial conditions for massive star formation.

In this paper we study the physical conditions (temperature, column density, density) and  gas kinematics in the region around  G326.27-0.49 based on new mapping observations of the emission from $J$=3--2 transitions of CO (and \thCO) using APEX. The analysis also includes complementary archival maps of  $J$=2--1 transitions of $^{13}$CO observed within the SEDIGISM programme along with the mid-, far-infrared and sub-millimeter maps of dust continuum emission from the Spitzer, Herschel missions and Planck-ATLASGAL \citep{Csengeri2016}.

\section{Observations \& Datasets}

\subsection{Molecular Line Observations with LAsMA/APEX}

Spectra for $J$=3--2 rotational transitions of $^{12}$CO and \thCO\ were
observed with the 7-pixel receiver, the Large APEX sub-Millimetre Array (LAsMA)  
on October 10, 2021 in very good weather conditions (precipitable water vapor, pwv  0.6 mm) 
using the APEX\footnote{APEX, the Atacama Pathfinder Experiment is a
collaboration between the Max-Planck-Institut für Radioastronomie,
Onsala Space Observatory (OSO), and the European Southern Observatory
(ESO).} telescope \citep{guesten2006}. This heterodyne spectrometer
allows simultaneous observations of the two CO isotopologues in the upper
(\twCO~at 345.7959899\,GHz) and lower (\thCO, 330.5879653\,GHz) sideband
of the receiver. The LAsMA backends are fast Fourier transform
spectrometers (FFTS) with 4\,GHz bandwidth \citep{Klein2012}
and a native spectral resolution of 61 kHz.  The array is configured in
a hexagonal arrangement around a central pixel with a spacing of about
two beam widths between the pixels. The main beam size of the APEX
observations is $\theta_{\rm mb}$ = 18\farcs2 at 345.8 GHz.

A 700\arcsec $\times$ 600\arcsec\ map centered at R.A.=15$^h$47$^m$10.8$^s$ Dec=-55\arcdeg
11\arcmin 12\arcsec\ (J2000) was observed at 345\,GHz. The observations were performed by
scanning in the total power on-the-fly mode, with a spacing of 9\arcsec\ between rows,
while oversampling at 6\arcsec\ in scanning direction, R.A..  The map data were calibrated
against a sky reference position at (RA 15$^h$ 54$^{\rm m}$ 18.5$^{\rm s}$, Dec -55\arcdeg
40\arcmin 28\arcsec).  

For the APEX observations, a  first order baseline was removed, and a
main beam efficiency $\eta_{\rm mb}$ = 0.68 was used to convert the
antenna temperature to main beam temperature. The reduced spectra have
been binned to a spectral resolution of  0.5 km/s. The final CO data
cubes (pixel size 9\farcs1) reveal an rms of 0.25\,K.

We have additionally performed deep integrations (60-80\,mK rms at a resolution of 0.5\,\kms) in the $J$=4--3 transition of HCO$^+$ and HCN (on Oct 11 \& 14, 2021) towards the four prominent cores in the region. These observations were performed in chopped mode, with a wobbler throw of 200\arcsec\ in R.A., since the emission is compact in these high-density tracers, and only detected in the central pixel of LAsMA.

\subsection{$^{13}$CO(2--1) Emission from SEDIGISM Survey}

We have used \thCO(2--1) archival data of the region obtained as part of
the programme Structure, Excitation and Dynamics of the Inner Galactic Interstellar Medium \citep[SEDIGISM;][]{schuller2021} for comparison with the higher frequency APEX observations. The data with a spatial resolution of 31\farcs7 was resampled 
to a final datacube with a pixel size of 9\farcs1 and a spectral resolution of 0.5\,\kms\ resulting in a typical rms of 0.26\,K.

\subsection{Archival Infrared and Sub-millimeter Continuum Data}

We have used infrared continuum emission maps of the region observed
with Spitzer/IRAC, Spitzer/MIPS, Herschel/PACS, Herschel/SPIRE and
APEX/Laboca. The 3.6, 4.5, 5.8 and 8.0\,\micron\ maps at $\sim 2$\arcsec\ resolution
were observed as part of the GLIMPSE programme \citep{benjamin2003} and the 24\,\micron\
data were obtained as part of the MIPSGAL programme
\citep{carey2009,gutermuth2015}, both observed using Spitzer. The
far-infrared emission maps at 70, 160, 250, 350, and 500\,\micron\ were
observed with beam-sizes of 5\farcs6, 10\farcs7, 17\farcs6, 23\farcs9 and 35\farcs2 respectively, 
as part of the Hi-GAL key programme of the Herschel \citep{Molinari2010}. The 870\,\micron\ map 
(with a beamsize of 19\farcs2) was generated from a combination of the 870\,\micron\ data  
observed as part of the ATLASGAL survey programme \citep{schuller2009} and the
353\,GHz {\em Planck}/HFI observations \citep{Csengeri2016}.

\section{Overview of the G326 Region}
\subsection{Morphology}

\begin{figure*}
\includegraphics[width=18.0cm]{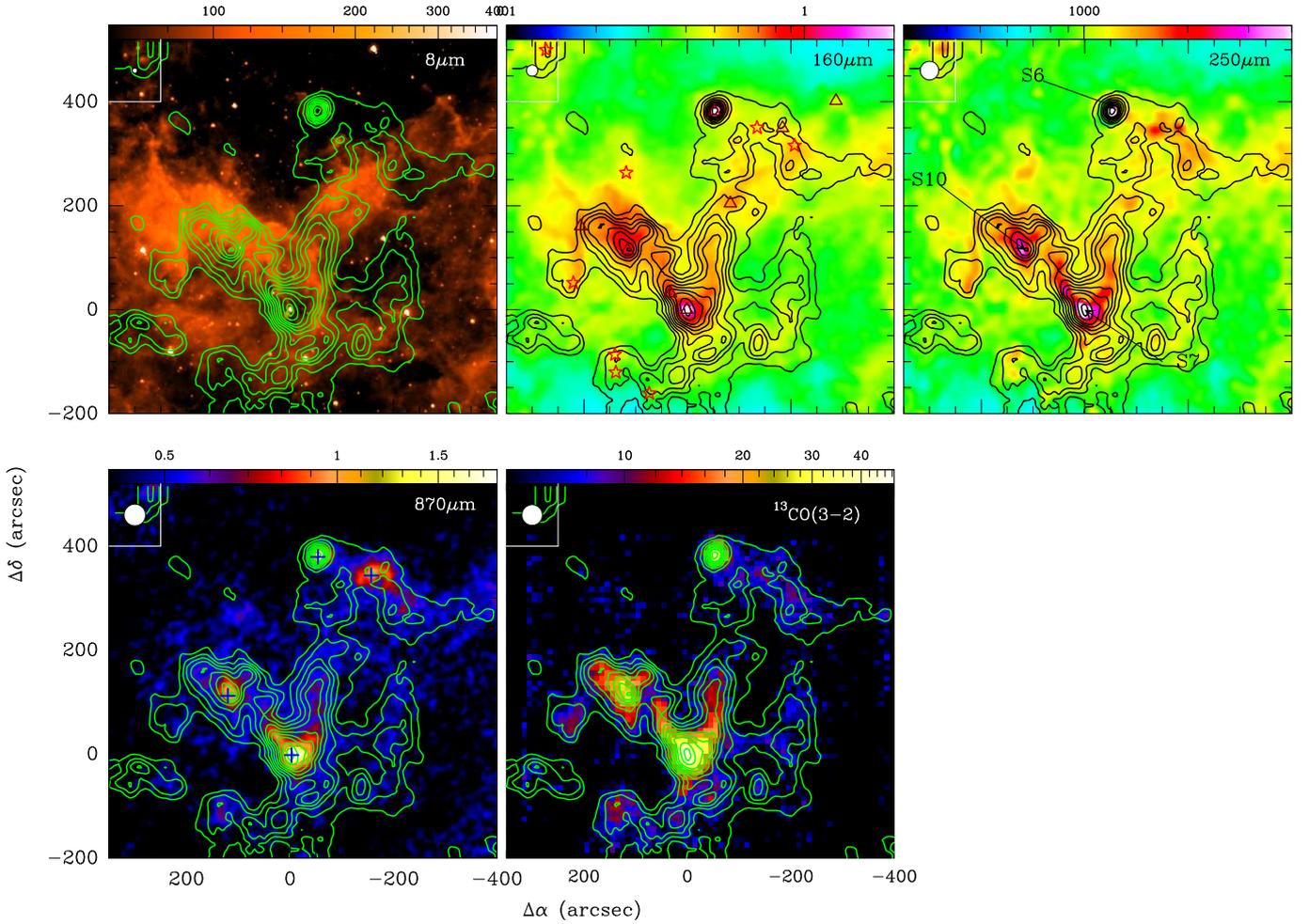}
\caption{Maps of continuum emission at 8, 160, 250, and
870\,\micron\ in color. The contours correspond to the CO(3--2) emission
integrated between -75 to -54\,\kms\ and have values of 20, 25, 30, 40
to 130 (in steps of 10) K\,\kms.  The color scale is shown above each
panel, the unit of flux density is MJy\,sr$^{-1}$ at 8\,\micron, Jy\,pixel$^{-1}$ at 160\,\micron, MJy\,sr$^{-1}$ at 250\,\micron, and Jy\,beam$^{-1}$ at 870\,\micron\ for a beam size of 21\,\arcsec. The \thCO(3--2) is integrated between -75 to -54\,\kms\ and expressed in units of K\,\kms. The
coordinates are shown as offsets in arc seconds relative to the center
RA:15$^{\rm h }$47$^{\rm m}$10.8$^{\rm s}$ Dec:-55\arcdeg 11\arcmin 12\arcsec.
Sources identified in the Hi-GAL catalog are marked on the 160\,\micron\
map, the triangles correspond to the 70\,\micron-bright sources,
whereas the star-shaped symbols show 70\,\micron-dark sources
\citep{elia2017}. The point sources identified in the ATLASGAL+{\em
Planck} maps are marked on the 870\,\micron\ map as  '+'. Location of the three main sources identified in this region S6, S7 and S10 are shown on 250\,\micron.
\label{fig_contover}}
\end{figure*}

We study the morphology of the G326 region by comparing the dust
continuum emission at 8, 160, 250 and 870\,\micron\ with the
velocity-integrated (-75 to -54\,\kms) CO(3--2) emission
(Fig.\,\ref{fig_contover}). The 8\,\micron\ map which primarily
traces emission from the transiently FUV-heated Polycyclic Aromatic
Hydrocarbon (PAH) molecules, reveals diffuse emission throughout the region 
where CO(3--2) is detected. With the exception of the south-eastern CO(3--2)
peak, S10, no other CO(3--2) peak shows any emission in the
8\,\micron\ image apart from some point-like sources. The far-infrared
continuum emission however reveals embedded bright sources as well as
more extended emission close to the CO(3--2) peaks. The 
mapped area includes two actively star-forming regions: a southern V-shaped ridge with two prominent continuum sources, one at the tip of the V and the other at the end of
the eastern-arm of the V;  a northern east-west extended filament-like
structure with a strong and compact source to the east and a more
extended structure to the west.  The two regions are connected by
comparatively faintly emitting molecular material. Based
on the band-merged Hi-GAL catalog \citep{elia2017}, a total of sixteen far-infrared sources are
identified in this region (Table\,\ref{tab_fir}). We have re-calculated the mass and bolometric 
luminosities of the far-infrared sources from the values given by \citet{elia2017} using the revised distance estimates provided by CO observations by \citet{duarte-cabral2021}.
The southern ridge hosts a total of 4 Hi-GAL sources that are detected from
70--500\,\micron, only two of which (S7 and S10) are also identified at
870\,\micron.  Sources S8 and S14 being fainter are not detected at 870\,\micron. 
The northern ridge also hosts a total of four Hi-GAL
sources, only two of which are detected at 70\,\micron\ and the source
to the east is the brightest in both continuum and line emission. Based on the 
distances of the northern and southern ridges as estimated from the velocity information
in the CO observations \citep{duarte-cabral2021}, and from present
observations as detailed later, it is likely that the ridges although at
slightly different distances are interacting.  In addition to the
protostellar sources identified in the Hi-GAL data, the southern ridge
hosts three dark clouds \citep{peretto2016} and the two EGOs likely to be MYSOs \citep{cyganowski2008} 
in the region coincide with the brightest far-infrared and CO(3--2) emission peaks.

\begin{table*}
\caption{Far-infrared sources in the  G326 region detected by the band-merged Hi-GAL catalogue \citep{elia2017}. The 870\,\micron\ flux is from the ATLASGAL catalogue. \citet{elia2017} estimated dust temperature ($T_{\rm dust}$; Col. 9), mass and luminosity of sources by fitting the SEDs with greybodies. 
The mass (Col. 11) and bolometric luminosities (Col. 12) have been re-calculated using the revised
distance estimates (Col. 10)
\label{tab_fir}}
\begin{tabular}{rcrrrrrrrrrr}
\hline
\hline
\scriptsize
Source & Coordinates & $F_{70}$ & $F_{160}$ & $F_{250}$ & $F_{350}$ &
$F_{500}$ & $F_{870}$ & $T_{\rm dust}$ & D$^a$ & Mass & $L_{\rm bol}$\\
& (h:m:s d:m:s) & Jy & Jy & Jy & Jy & Jy & Jy  & K & kpc & \msun & \lsun\\ 
1 & 2 & 3 & 4 & 5 & 6 & 7 & 8 & 9 & 10 & 11 & 12\\
\hline
S1 & 15:46:37.34  -55:04:29.7& 3.5$\pm$0.3 & 5.7$\pm$0.4 & 4.2$\pm$0.3 & 3.6$\pm$0.3 & \ldots & \ldots & 17.4$\pm$1.2 & 4.1
&24.4$\pm$24.4 & 166.1 \\
S2 & 15:46:46.67  -55:05:55.5 & \ldots & 6.6$\pm$0.5 & 12.6$\pm$1.0 & 19.9$\pm$3.6 & 6.0$\pm$2.5 & \ldots & 11.4$\pm$0.7 &
 4.1 & 404.7$\pm$141.9 & 83.8\\
S3 & 15:46:49.37  -55:05:21.7 & 0.7$\pm$0.1 & 3.4$\pm$0.3& 	8.2$\pm$0.7 & 33.7$\pm$2.4 & \ldots & \ldots & 10.0$\pm$0.2 &
4.1 & 745.3$\pm$154.1 & 104.3\\
S4 & 15:46:55.10  -55:05:20.9 & \ldots & 3.0$\pm$0.2 & 14.9$\pm$0.8 & 32.9$\pm$2.8 & 18.3$\pm$1.1 & 0.69$\pm$0.14 & 9.2$\pm$0.2 &
4.1 & 1564$\pm$187 & 79.1\\
S5 & 15:47:01.14 -55:07:46.8 & 1.5$\pm$0.5  & 10.7$\pm$1.3 &  8.3$\pm$0.8  & 3.8$\pm$0.3  & \ldots  & \ldots & 19.9$\pm$1.6 & 3.6 & 25.7$\pm$25.5 & 159.8\\
S6 & 15:47:04.72  -55:04:49.8 & 45.2$\pm$0.5 & 96.6$\pm$1.8 & 102.3$\pm$1.9 & 46.7$\pm$0.8 & 19.0$\pm$0.4 & 1.99 &
22.0$\pm$0.2 & 4.1 & 214.9$\pm$8.9 & 2325.8\\
S7 & 15:47:10.90 -55:11:11.3 & 81.4$\pm$3.9 & 121.9$\pm$2.4& 125.3$\pm$4.1 & 49.4$\pm$1.9 & 32.1$\pm$1.5 &3.6$\pm$0.0 &
19.3$\pm$0.3 & 3.6 & 315.9$\pm$21.1 & 2969.6\\
S8 & 15:47:14.60 -55:10:39.5 & 3.3$\pm$0.2  & 13.1$\pm$1.3 &  23.9$\pm$1.8 & 11.1$\pm$1.0 & 9.2$\pm$1.9 & \ldots &
13.8$\pm$0.7 & 3.6 & 199.2$\pm$54.9 & 233.4\\
S9 & 15:47:19.41 -55:13:52.0 & \ldots & 3.4$\pm$1.7	 & 24.8$\pm$2.0 & 5.3$\pm$0.5 &	2.6$\pm$0.4 & \ldots & 13.6$\pm$1.3 &
3.6 & 109.3$\pm$108.5 & 67.4 \\
S10 & 15:47:24.31 -55:09:16.9 & 3.0$\pm$0.2  & 10.6$\pm$0.7 &  46.8$\pm$2.1 & 31.7$\pm$1.8 & 12.8$\pm$0.6 & 1.5$\pm$0.0 &
11.4$\pm$0.2 & 3.6 & 718.4$\pm$67.5 & 233.5\\
S11 & 15:47:24.59  -55:06:47.6 &\ldots &	12.6$\pm$1.6 &	23.8$\pm$1.8 & 7.7$\pm$0.4 & 6.1$\pm$0.5 & \ldots &14.2$\pm$1.0
&3.6&180.9$\pm$179.5 &141.1\\
S12 & 15:47:26.96 -55:13:11.7 & \ldots & 1.9$\pm$0.4 & 6.1$\pm$1.0 & 16.2$\pm$1.3 & \ldots & \ldots & 9.2$\pm$0.5 & 3.6 &
609.4$\pm$156.4 & 47.8\\
S13 & 15:47:27.18 -55:12:37.6 &\ldots & 3.4$\pm$0.8 & 16.4$\pm$1.8 & 22.5$\pm$1.8& 9.5$\pm$0.5 & \ldots & 10.7$\pm$0.5 & 3.6&
405.7$\pm$72.4 & 55.3\\
S14 & 15:47:34.90 -55:08:30.9 & 4.3$\pm$0.4  & 7.0$\pm$1.0  &   6.2$\pm$0.7 & 3.7$\pm$1.2  & \ldots  & \ldots & 16.4$\pm$1.8 &
3.6 &31.9$\pm$31.7 & 156.7\\
S15 & 15:47:36.54 -55:10:19.6 & \ldots & 7.8$\pm$1.6 & 21.6$\pm$2.1&	12.7$\pm$1.2 & 4.7$\pm$0.4 & \ldots  & 13.2$\pm$0.8
&3.6&180.9$\pm$179.5 & 90.3 \\
S16 & 15:47:42.59  -55:02:51.5 & \ldots & 6.9$\pm$0.6 & 15.6$\pm$1.8 & 11.2$\pm$1.4 & 6.8$\pm$0.7 & \ldots & 11.9$\pm$0.5 & 4.1 &
329.4$\pm$70.9 & 86.7\\
\hline
\hline
\end{tabular}
\end{table*}

\subsection{Kinematics}

\begin{figure*}
\includegraphics[width=0.95\textwidth]{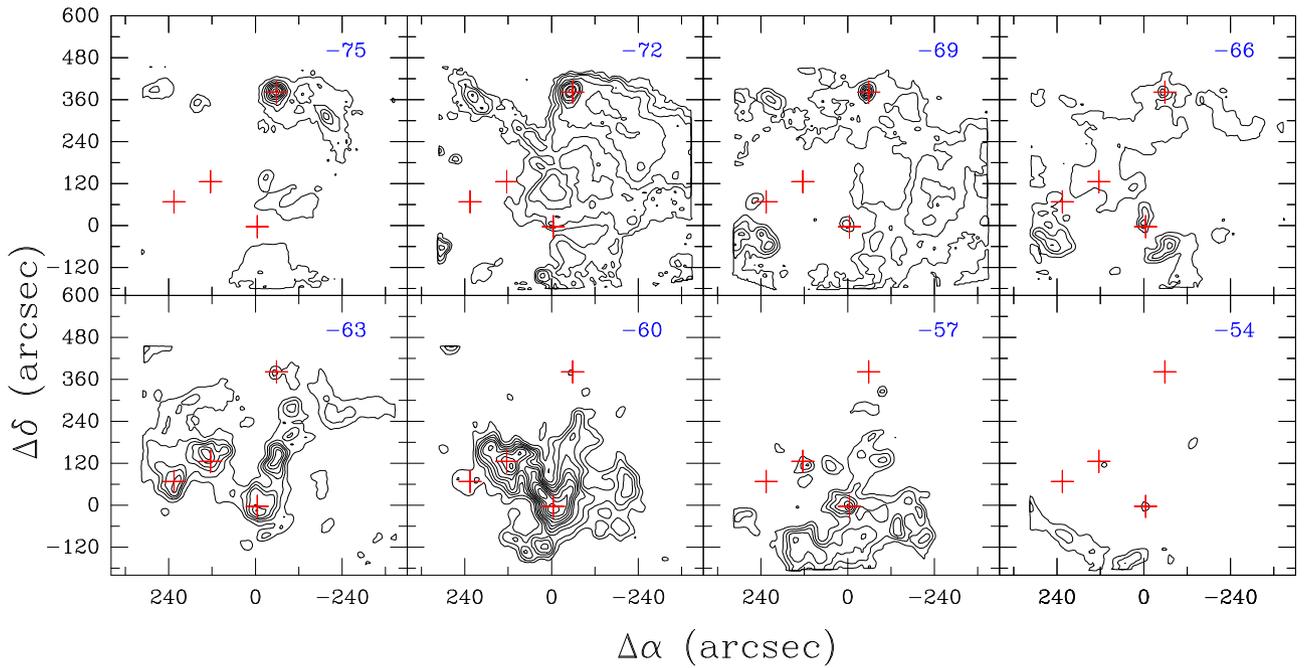}

\caption{Channel Map of CO(3--2) with width of each channel being 3\,\kms.  Contour levels are : (1,2,\ldots 7, 8.5, 9.5\,K) for -75\,\kms, (1, 1.5, 2, 3,\ldots 7, 8.5, 9.5, 11.5\,K) for -72\,\kms, (1, 2, 3.5, 5, 6.5, 8.5, 9.9\,K) for -69, -66, -63\,\kms\ and (2, 3.5, 5, 6.5, 8.5, 10, 12, \ldots 18\,K) for -60, -57 and -54\,\kms.  The red '+' mark the position of the sources S6, S7, S10 and S15. The axes mark offsets in arcseconds relative to the center 15$^{\rm h }$47$^{\rm m}$10.8$^{\rm s}$ -55\arcdeg 11\arcmin 12\arcsec\ (J2000) \label{fig_chanmap_co32}}
\end{figure*}

\begin{figure*}
\includegraphics[width=0.95\textwidth]{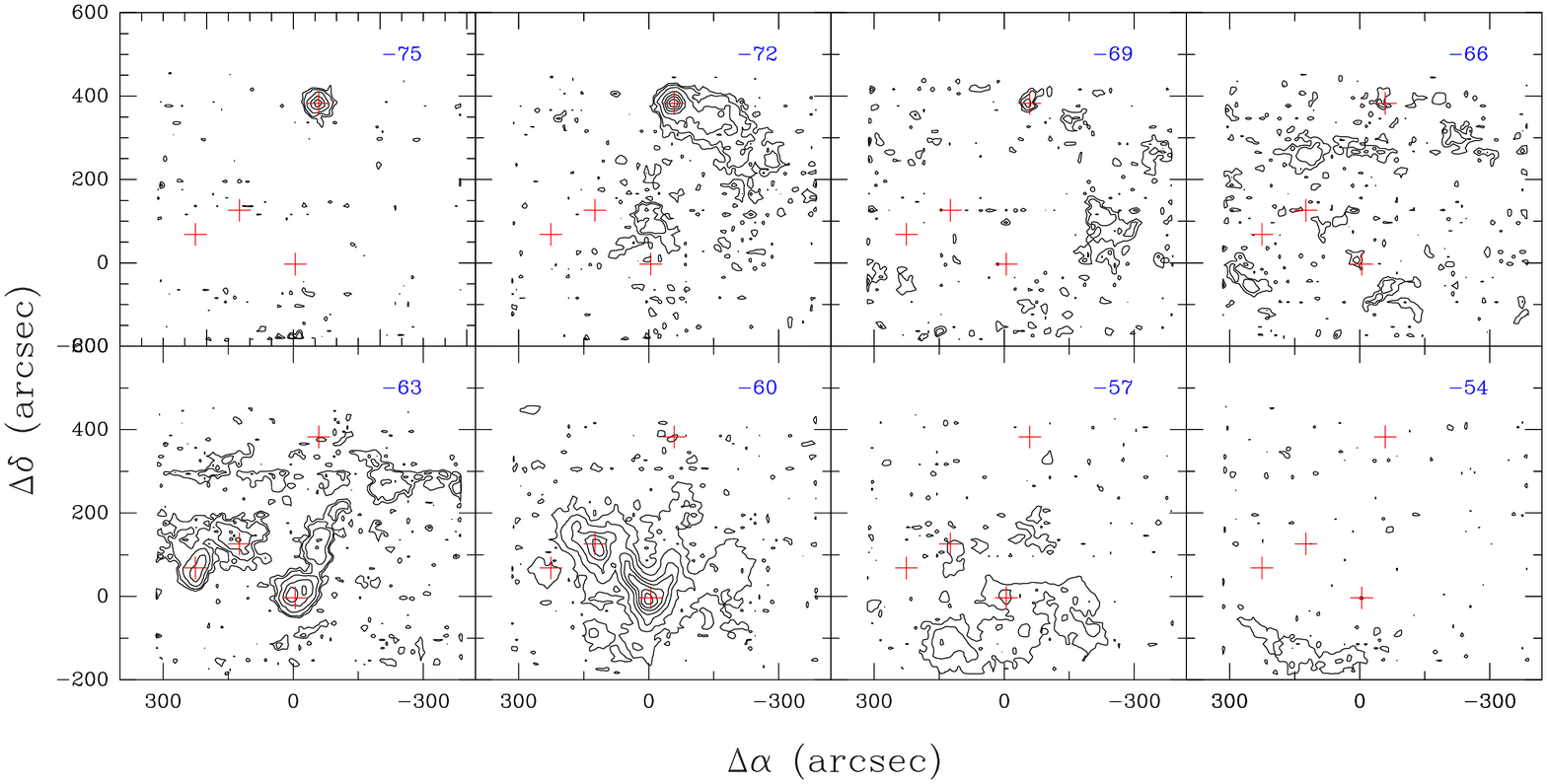}
\caption{Channel Maps of \thCO(3--2), width of each channel being 3\,\kms. Contour levels for \thCO(3--2) are 0.3, 0.5, 1 ,2,\ldots 7 K\,\kms. Rest of the details same as in Fig.\,\ref{fig_chanmap_co32} \label{fig_chanmap_13co32}}
\end{figure*}

The velocity-channel maps of CO(3--2) and \thCO(3--2) emission from the
region were constructed with data smoothed to a resolution of 3\,\kms\ for the ease of display (Fig.\ref{fig_chanmap_co32}, \ref{fig_chanmap_13co32}). The channel maps
show that the region consists of two main emission features centered at
-72\,\kms\ and -60\,\kms\ that lie to the north and south respectively
of the region. At -72\,\kms\ we also detect a faint CO(3--2) emission
that spatially overlaps with the -60\,\kms\ cloud, this diffuse emission
is likely fainter than the detection limit in the \thCO(3--2) map. The intensity peaks centered at -63\,\kms\ coincide with the Hi-GAL sources S7, S10 and S15, while the emission from S6 peaks in the -75\,\kms\ bin. The channel maps indicate spatial coincidence of the multiple velocity components at a few locations, however the signature of physical interaction between the components, if any, is not entirely obvious.  We have explored possible interaction between the
molecular features at different velocity intervals using the
position-velocity ($p$--$\upsilon$) diagrams in Sec. 5.2. 

\subsection{G326 as a Hub-Filament-System}

Based on Herschel continuum images the star forming region located at G326.27-0.49 was
identified as a Hub-filament-System with the source S7 being at the hub \citep{kumar2020}.
While, the continuum images have high sensitivity for the detection of structural details,
the lack of velocity information leaves room for the possibility of overlap of emission
features not belonging to the same system but being located along the same line of sight.
In the case of the region around G326, the CO(3--2) channel map clearly shows filamentary
structure particularly at the velocities of -63 and -60\,\kms. In order to explore whether
the structures seen in the CO(3--2) velocity-integrated and dust continuum maps are
coherent in velocity, we have identified filaments in the region by applying the python
package FilFinder \citep{Koch2015} on both the CO(3–2) and \thCO(3--2) datacubes at a resolution of 1\,\kms.
Details of the parameters used in the analysis are presented in Appendix A. For the
-60\,\kms\ cloud component, the overlay of filaments identified in the velocity slices of the
datacube between -63 to -55\,\kms\ clearly shows that the 
structure seen in the integrated intensity maps of the region arises from 
gas that is coherent in velocity as well (Fig.\ref{fig_filfinder}). There are offsets between
the spines of the filaments identified in the CO(3--2) and \thCO(3--2) data, however for the 
purpose of verification of the velocity coherence of the structures this
is not significant. We note a  number of such filaments overlap and
interact forming a spiral structure at the location of the 'hub' that
hosts the high-mass protostar S7.  This provides further evidence that the region located at G326.27-0.49 hosts a velocity-coherent hub-filament-system with high-mass star formation activity in the hub. We further emphasize that at the resolution of this study, the elongated structures that we identify are likely to be a bundle of filaments with a velocity gradient perpendicular to the axis as well.

\section{Column density of molecular gas in G326}
\subsection{Estimates from dust continuum emission}

In order to obtain an overview of the distribution of cold molecular gas that is essentially the reservoir for material forming stars, we first use the far-infrared emission maps between 160 to 350\,\micron\ and at 870\,\micron\ to obtain the distribution of column density of the cold dust in the region. For this all the maps were smoothed to a common resolution of 24\arcsec\ and regridded to the same pixel size and then a pixel-by-pixel fitting of the intensities was performed using a modified blackbody function of the form

\begin{equation}
F_{\nu} = \Omega_{\rm pix} B_\nu (T_{\rm d}) (1-e^{-\tau_\nu}),
\end{equation}

where $\Omega_{\rm pix}$ is the solid angle for each pixel and $B_\nu(T_{\rm d})$ is
the blackbody function at a dust temperature of $T_{\rm d}$. The optical depth
($\tau_\nu$) is defined as 

\begin{equation}
\tau_\nu = \mu m_{\rm H} \kappa_\nu N({\rm H_2})
\end{equation}

where $\mu$ is the mean weight of the molecular gas taken to be 2.86 assuming that the gas is 70\% molecular hydrogen by mass \citep{ward-thompson2010}, $m_{\rm H}$ is the mass of hydrogen atom, $\kappa_\nu$ is the dust opacity, and $N$(H$_2$) is the column density estimated following \citet{ward-thompson2010}

\begin{equation}
\kappa_\nu = 0.1\left(\frac{\nu}{1000\,{\rm GHz}}\right)^\beta
\end{equation}

where $\nu$ is the frequency, $\beta$ is the dust emissivity index assumed to be 2 in our analysis. 

\begin{figure*}
\includegraphics[width=0.48\textwidth]{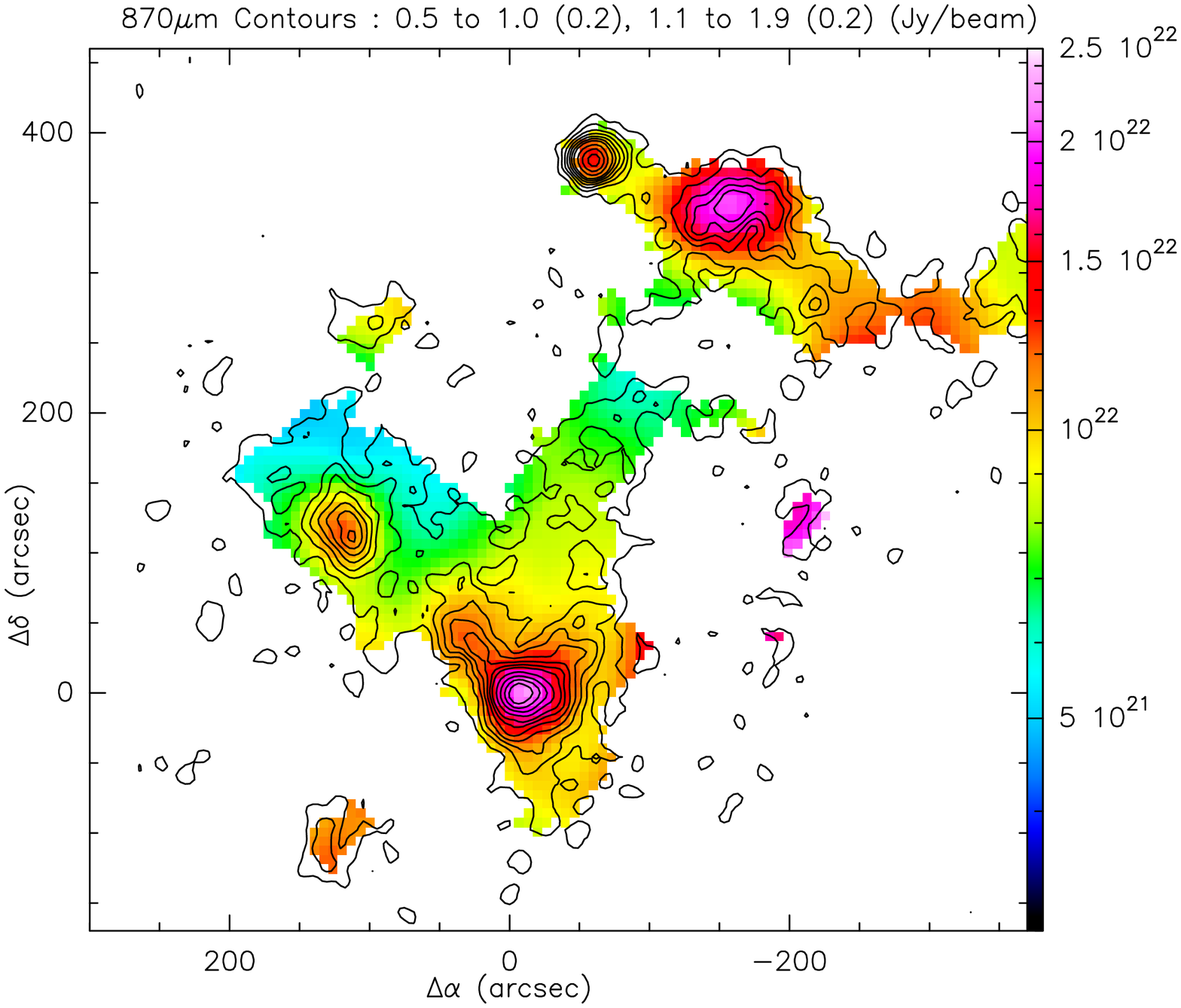}
\includegraphics[width=0.45\textwidth]{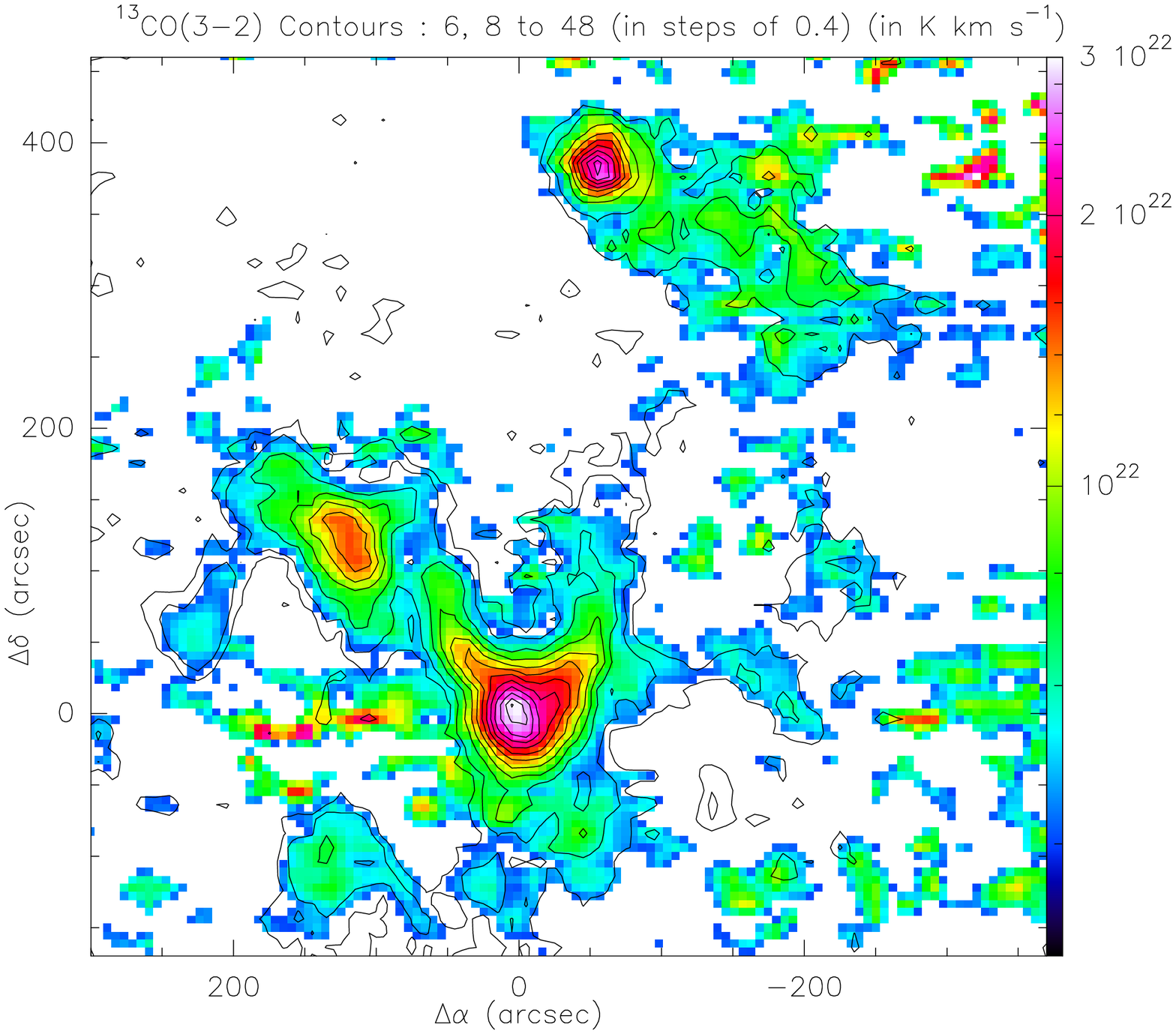}
\caption{Maps of $N$(H$_2$) column density in the region obtained from
dust and gas emission both at a resolution of 24\arcsec. {\em
Left:}Obtained by pixel-by-pixel fitting of 160, 250, 350 and
870\,\micron\ continuum emission with a grey-body function for a dust
emissivity exponent ($\beta$) of 2.0 shown in color. The Planck-ATLASGAL
map of 870\,\micron\ continuum emission is plotted as contours with
levels mentioned above the panel and {\em Right:} Derived from \twCO\ and
\thCO(3--2)  emission maps. Contours correspond to \thCO(3--2) emission
integrated between -75 to -54\,\kms, contour levels shown above the
panel. 
\label{fig_tempcdens}}
\end{figure*}

We restricted the fit to flux densities at 160, 250, 350 and
870\,\micron\ because (a) 70\,\micron\ traces warm dust and would thus
necessitate the use of two dust components with different temperatures
during the fitting and (b) inclusion of 500\,\micron\ would have
necessitated degrading the column density and dust temperature maps to a
resolution of  37\arcsec, in contrast to the 24\arcsec\ used currently.
Figure\,\ref{fig_tempcdens} shows the distribution of the column density of
cold gas as derived from the dust emission maps using the above method. The
dust temperature was found to vary between between 16 to 20\,K
(Fig.\,\ref{fig_dtemp}) since only the longer wavelength emission tracing the cold dust component were used and the total column density $N(\rm H_2)$ was found to vary between 6$\times 10^{21}$--2.3$\times 10^{22}$\,\cmsq.

\subsection{Estimates from CO and $^{13}$CO(3--2) emission}

We have used the \twCO\ and \thCO(3--2) emission maps to derive the $N({\rm H_2})$ map for the region using the formalism outlined by \citet{Tiwari2021}. We first estimated the excitation temperature ($T_{\rm ex}$) at every pixel of the CO(3--2) map by assuming the \twCO\ to be optically thick under LTE using 

\begin{equation}
T_{\rm ex} = 16.6\left[\ln\left(1+\frac{16.6}{{\rm T_{\rm mb}(^{12}CO)}}\right)\right]^{-1}\,K.    
\end{equation}

where $T_{\rm mb}$ is the main-beam brightness temperature. The $T_{\rm ex}$ thus derived
from the region varies between 15 to 35\,K, with values near the sources S7 and S10 being
closer to 30\,K (Appendix B) and near the source S6 being $\sim 20$\,K
(Fig.\,\ref{fig_tex}). Inspection of the \twCO(3--2) spectra reveal strong self-absorption
features particularly towards the source S6, and to some extent towards S7 and S10. Thus
the excitation temperatures estimated in this way are liable to be under-estimated around
these positions, leading to a possible over-estimate of the column density. We also point out that the $T_{\rm dust}$ of 16--20\,K derived 
from continuum emission from the cold dust traced at wavelengths longer than 160\,\micron\ 
is slightly lower than but consistent with the $T_{\rm ex}$ estimated here.

 Using the excitation temperature map so generated, assuming the excitation temperatures of CO and \thCO\ to be similar (an assumption that could lead to an underestimation of the \thCO(3--2) optical depth for densities, $n_{\rm H_2}<10^4$\,\cmcub\ and the \thCO(3--2) map  of intensity integrated between -75 to -54\,\kms, we have estimated $N$(\thCO) using:

\begin{eqnarray}
N(^{13}{\rm CO}) & = &5.29\times 10^{12}\left(T_{\rm ex}+0.88\right)\times \exp \left(\frac{31.7}{T_{\rm ex}}\right) \nonumber\\
&\times & \left(\frac{\tau_{13}}{1-\exp(-\tau_{13})}\right)\displaystyle \int {T_{\rm mb}(^{13}{\rm CO})d\upsilon}~~{\rm cm^{-2}}
\end{eqnarray}

where the numerical factors are as described by \citet{Tiwari2021} and $T_{\rm mb}$(\thCO) is the peak
main-beam temperature of the \thCO(3--2) spectrum and the \thCO\ optical depth, $\tau_{13}$ is calculated using

\begin{equation}
\tau_{13} = -\ln\left[1-\frac{T_{\rm MB}(^{13}{\rm CO})}{15.87}\times\left(\frac{1}{\exp(15.87/T_{\rm ex})-1}-0.003\right)^{-1}\right]
\end{equation}

The estimated $\tau_{13}$ varies between 0.4 to 1.1, with the region
near S7 and S10 showing values of 0.5--0.6 and the source S6 showing
values between 0.8--1.0 (Fig.\,\ref{fig_tex}). We consider
[H$_2$]/[\twCO]=10$^4$ and  [\twCO]/[\thCO]=50 to obtain the  column
density map of the region following the method described. The column
density map, originally at a resolution of 20\arcsec, was smoothed to a
resolution of 24\arcsec\ to match the $N$(H$_2$) map obtained from dust emission (Fig.\,\ref{fig_tempcdens}).  Although the observed CO and \thCO(3--2) emission are evidently more extended than the observed dust continuum emission at 870\,\micron, the values of column densities derived from the two approaches, particularly at the location of the three main sources S6, S7 and S10 identified in this region, are consistent within a factor of 1.5. We notice a second peak in the northern cloud only in the $N$(H$_2$) map
generated from the dust continuum that is also present in the continuum images at 250 and 870\,\micron\ (Fig.\,\ref{fig_contover}. The \thCO(2--1) data shows some enhanced emission at the position of this second peak, but the $J$=3--2 maps detect only diffuse emission suggesting the source to be a cold clump.



\section{Analysis of the 3-dimensional structure}

\subsection{Molecular clumps and their kinematic properties}

Presence of outflows and far-infrared sources indicate definite star
formation activity in the north and south filaments lying in the region
around G326 studied here. In order to understand the potential of the
HFSs as sites of ongoing and  future star formation, we
explore the kinematic stability of the molecular clumps and 
the region as a whole. For this we have used the dendrogram-based
structure analysis tool {\em astrodendro} \citep{2008ApJ...679.1338R} to
identify molecular clumps in both the CO(3--2) and the $^{13}$CO(3--2)
emission. While the CO(3--2) being brighter, identifies more clumps and
the filamentary structures in the region clearly, the lines being
optically thick and self-absorbed at many positions the masses of clumps
are not reliable. Hence we use the results of dendrogram analysis of the
CO(3--2) map for the study of the large-scale structure of the region
(Sec. 5.2), and use the \thCO(3--2) map to derive properties of the smaller scale structure (clumps).

The dendrogram technique decomposes the hierarchical structure of a
molecular cloud in three dimensional data cubes into a range of scales
known as trunks, branches and leaves. The structure of the dendrogram
depends on the local maxima in the data, which determine the top level
of the dendrogram that is referred to as leaves. Leaves are the set of
isosurfaces that contain a single local maximum. Branches are the
sub-structures within a dendrogram tree that contain leaves and other
branches. Trunks are defined as structures that have no parent structure
and form the base of a dendrogram tree. For the dendrogram analysis, we
created a sub-cube of $^{13}$CO(3--2) data in the velocity range -75 to
-54~\kms. The selected velocity range covers the emission from the
entire HFS and the secondary cloud. An intensity threshold (min$\_$value) of 5$\sigma$ where
$\sigma$ is rms noise is chosen to filter out the structures with low
signal to noise ratio. We have also set the min$\_$delta parameter as
2$\sigma$ such that a leaf is considered independent if and only if its
peak intensity is above 2$\sigma$ compared to the peak of the
neighbouring leaf or branch. In addition, the min$\_$npix parameter, the
minimum number of pixels needed for a leaf to be considered an
independent entity is set such that the area of the identified structure
is at least 1.5 times the area of the beam. Figures\,\ref{fig_dendro}
and \ref{fig_dendro2} show the results of the dendrogram analysis on
the \thCO(3--2) and CO(3--2) emission maps respectively. We identify a
total of 39 and 25 leaves respectively in the dendrogram tree
constructed from the CO(3--2) and \thCO(3--2) maps (denoted as CL\# for the rest of the text).  Since the
\thCO(3--2) lines are optically thin, we use the clumps obtained from
the dendrogram analysis of the \thCO(3--2) map to derive the masses and
kinematic stability of the structures in the region.

Table\,\ref{tab_dendro} presents the location, size, velocity, linewidth, column density and mass  of the
clumps identified in the \thCO(3--2) map using {\em astrodendro}. The deconvolved size (diameter, $\sigma_{\rm
eff}$) of the clumps are derived first by  subtracting the beam width ($\theta_{\rm b}$=20\arcsec) in
quadrature following the Equation (7). The radius ($R_{\rm eff})$ of each clump is obtained by using the
deconvolved size (angular diameter) and an assumed distance of 4\,kpc. We identify clumps with sizes ranging
between 0.1 to 0.46\,pc.  

\begin{center}

\begin{equation}
\sigma_{\rm eff}^2 = \sigma_{\rm meas}^2 - \left(\frac{\theta_{\rm b}}{8\ln 2}\right)^2
\end{equation}

\end{center}

We calculate the column density ($N_{\rm cl}$(H$_2$)) of each clump
by using Eq. (5), from the integrated \thCO(3--2) flux densities
estimated by {\em astrodendro} assuming an average excitation
temperature of 25\,K and an average \thCO\ optical depth ($\tau_{13}$)
of 0.5 (Table:\ref{tab_dendro}). As before we have assumed
[\twCO]/[\thCO]=50  and [H$_2$]/[\twCO]=10$^4$ \citep{SzHucs2014} for the calculation of the column density. In order to obtain a more
accurate mass estimate we have considered the area identified for the clumps and obtained
mean $T_{\rm ex}$ ($\tau_{13}$) of 15\,K (0.7), 28\,K (0.5) and 24\,K (0.4) for the sources
S6, S7 and S10 respectively. We note that the numerical factor to convert the \thCO(3--2) integrated intensities to column densities for all the relevant T$_{\rm ex}$ and $\tau_{13}$ values are within a factor of 1.6 of each other. It is expected that an additional uncertainty in the derived numbers is contributed by the self-absorption features in the \twCO(3--2) spectra leading to the under-estimate of the T$_{\rm ex}$.

We use the effective radius ($R_{\rm eff}$) and velocity dispersion ($\sigma_\upsilon$) of each clump to estimate the  virial parameters 

\begin{equation}
\alpha_{\rm vir} = 1.2\left(\frac{\sigma_\upsilon}{\rm km\,s^{-1}}\right)^2 \left(\frac{R}{\rm pc}\right)\left(\frac{\rm M}{10^3\,M_\odot}\right)^{-1} 
\end{equation}

For spherical and homogeneous density distribution, the virial parameter can also be written as $\alpha_{\rm vir} = \frac{2E_{\rm kin}}{|E_{\rm pot}|}$ \citep{BertoldiMckee1992}.  Since $\alpha_{\rm vir}$ is related to $E_{\rm kin}$/|$E_{\rm pot}$|, it can be used to study the kinematic stability of the clumps or cloud fragments. In the absence of pressure supporting the cloud $\alpha_{\rm vir}<1$ implies that the cloud is gravitationally unstable and collapsing, whereas $\alpha_{\rm vir}> 2$ means that the kinetic energy is higher than the gravitational energy and that the cloud is dissipating. A value of $\alpha_{\rm vir}$ between 1 and 2 is interpreted as an approximate equilibrium between the gravitational and kinetic energies. Additionally, a cloud undergoing gravitational collapse can also show $\alpha_{\rm vir}\sim 2$ as the rapid infall can manifest as a large velocity dispersion \citep{Kauffmann2013}.

For the twenty-five clumps identified in \thCO(3--2) emission in the
G326 region the $\alpha_{\rm vir}$ estimated are mostly larger than 4.
This is consistent with the outcome of a compilation of virial
parameters for structures ranging from entire molecular clouds
($>1$\,pc) to cores ($\ll$ pc) by \citet{Kauffmann2013} which shows a
large range of values of $\alpha_{\rm vir}$. Physically a value of
$\alpha_{\rm vir}$ exceeding 2 in the absence of magnetic fields implies
gas motions could prevent the structures from collapsing, thus implying
that the collapse towards star formation is a gradual process. For the
clumps CL11, CL15 and CL22 associated with the active star forming
far-infrared sources S6, S7 and S10, we obtain $\alpha_{\rm vir}$ of
2.7, 1.6 and 2.3 respectively, values that are smaller than the typical
values found in the region, but not indicative of being supercritical to
gravitational collapse. We note that a spatial resolution of 20\arcsec\
at a distance of 4\,kpc corresponds to a radius of 0.2\,pc. Hence it is
possible that the structures detected using the dendrogram analysis are
clumps (and not cores), only parts of which are collapsing to form the
high-mass protostellar objects that are detected in the far-infrared wavelengths and
harbor outflows.

\subsection{Position-Velocity Diagrams}

\begin{figure*}
\includegraphics[width=0.90\textwidth]{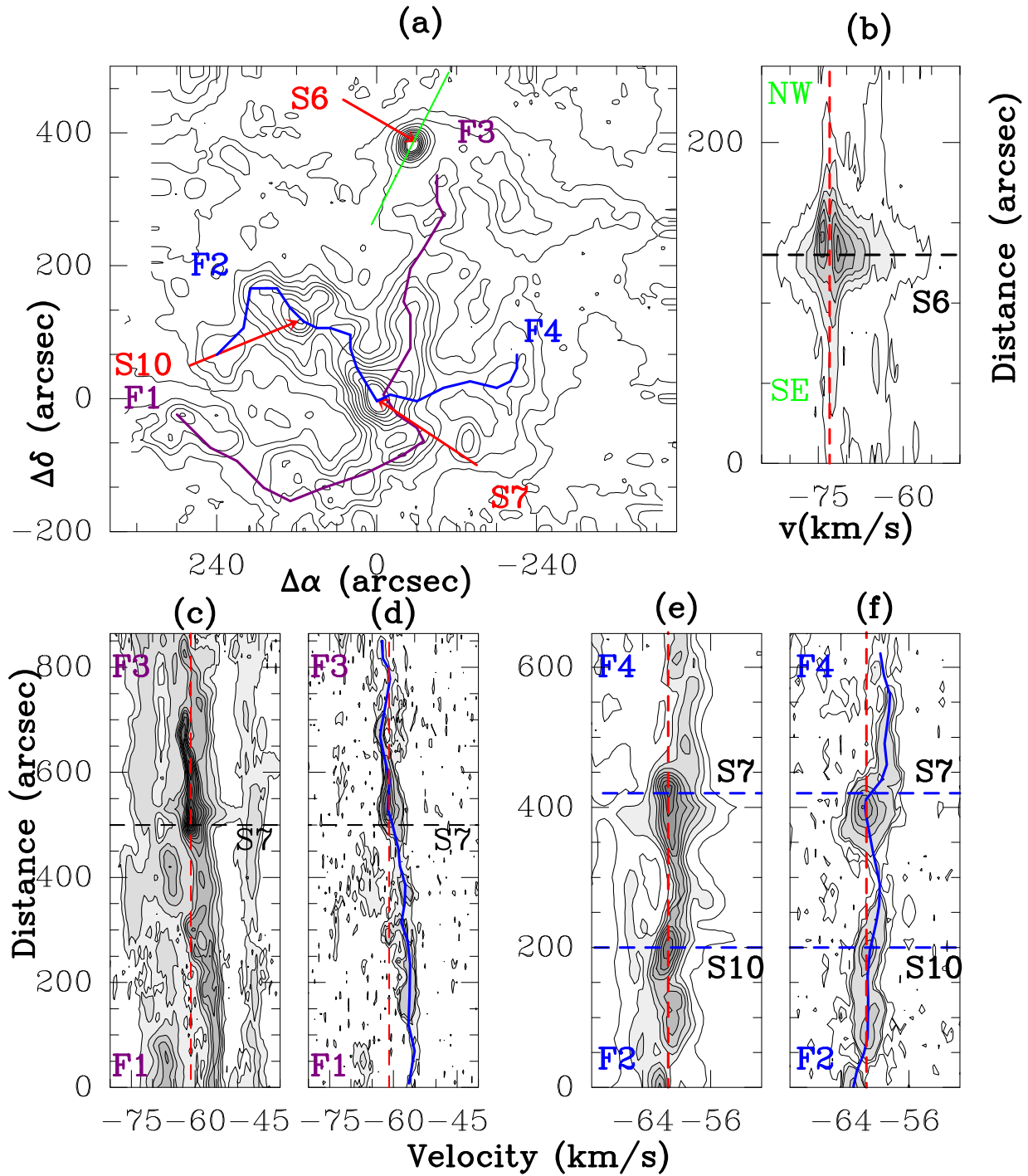}
\caption{{\em (a)} Intensity map of CO(3--2) integrated between -75 to -54\,\kms, shown with the 
directions along which position-velocity diagrams of CO(3-2) and \thCO(3--2) emission are obtained. For each $p$--$\upsilon$ diagram, the color of the text denoting the direction and endpoints are the same as the color of the cuts drawn on the intensity map. Panels (c) and (d) show the CO and \thCO\  $p$--$\upsilon$ diagrams respectively along the filaments F1 and F3. Panels (e) and (f) show the CO and \thCO\  $p$--$\upsilon$ diagrams respectively along the filaments F2 and F4. Contour levels: 
(a) 10 to 120 (in steps of 10) in K\,\kms, (b) 1 to 13\,K in steps of 2\,K (c) \& (f) 0.3, 1 to 22\,K in steps of 2\,K, (d) 0.3, 0.7, 1 to 9\,K in steps of 2\,K and (e) 1 to 27\,K in steps of 2.5\,K. Blue curves in panels (d) and (f) show the positions at which the velocity profiles plotted in Fig.\,\ref{fig_velprof} were measured.
\label{fig_pvdiag}}
\end{figure*}

Based on the features seen in the integrated intensity and channel maps
of CO (\thCO) (3--2) emission, and the structure identified using the
dendrogram-based analysis, we have selected directions across the
identified clumps and filaments to extract the position-velocity
($p$--$\upsilon$) maps along them (Fig.\,\ref{fig_pvdiag}).   The
directions F1F3 and F2F4 were chosen to create the $p$--$\upsilon$
diagrams so as to be able to analyse in detail the velocity structures along all the four filaments identified in the region. We have also marked the positions on the \thCO(3--2) $p$-$\upsilon$ diagrams where the velocities were measured to derive the profile (Fig.\,\ref{fig_velprof}) used to estimate the flow along the filaments. First, we note that there is an overall similarity of velocity
structures seen the CO and \thCO\ $p$-$\upsilon$ diagrams along the directions 
F1F3 and F2F4, although the CO(3--2) data detects the outflows from S7 and S10 and the two 
secondary clouds at velocities $\sim -70$\,\kms and $\sim -48$\,\kms\ in the F1F3 direction better.

The $p$--$\upsilon$ diagram along F1F3 traces the velocity distribution in the north-south direction, i.e., along the filaments F1 and F3 and shows a smooth velocity gradient of $\sim$0.6\,\kms\,pc$^{-1}$ with the primary component peaking at a velocity of -61\,\kms. F1 is clearly red-shifted with respect to the central peak, consistent with the results based on clump velocities. In addition, the $p$--$\upsilon$ diagram shows two secondary clouds at velocities $\sim -70$\,\kms and $\sim -48$\,\kms, respectively. Bridge-like structures are found to exist between these clouds and the primary cloud, mainly near the central hub, close to the outflow source S7. Such bridge features are often attributed to cloud-cloud collisions \citep[e.g.,][]{Haworth2015}.  Based on the moderate velocities of approach of the interacting structures we consider these clouds to be interacting with the main cloud rather than colliding with it.

The $p$--$\upsilon$ diagram along F2F4 traces the velocity distribution of the cloud along the filaments F2 and F4, in the NE-SW direction. The plot shows two prominent peaks around -61\,\kms. These correspond to the YSOs S7 and S10. Velocity of F2 is blue-shifted with respect to the peak velocity whereas F4 is red-shifted as seen from the dendrogram analysis. The overall velocity gradient is 0.4\,\kms\,pc$^{-1}$. The velocity gradients of 0.4--0.6\,\kms\,pc$^{-1}$ observed along both F1F2 and F3F4 are consistent with the typical velocity gradients of  0.2--2\,\kms\ as seen in filaments in low-mass star-forming regions \citep{kirk2013,peretto2014}. The $p$--$\upsilon$ diagram along a direction passing through the YSO S6 clearly shows the presence of an outflow extending over 20\,\kms, which appears to be the broadest of the three outflows detected in the region We discuss the properties of the outflows in Section 6. 

Finally, we note that the far-infrared continuum sources identified in F1 are in prestellar phase (no 70\,\micron\ emission) whereas those in F2 and F3 except one are in protostellar phase (detection in all five Herschel FIR bands) (Fig.\,\ref{fig_contover}). This implies that F1 is in an early evolutionary stage compared to filaments F2 and F3. Further, lack of bright continuum sources in F4 is suggestive of F4 also being in an early phase of formation compared to F2 and F3.

\subsection{Spiral Structure of and accretion flow in the filaments}

\begin{figure*}
\includegraphics[width=0.47\textwidth]{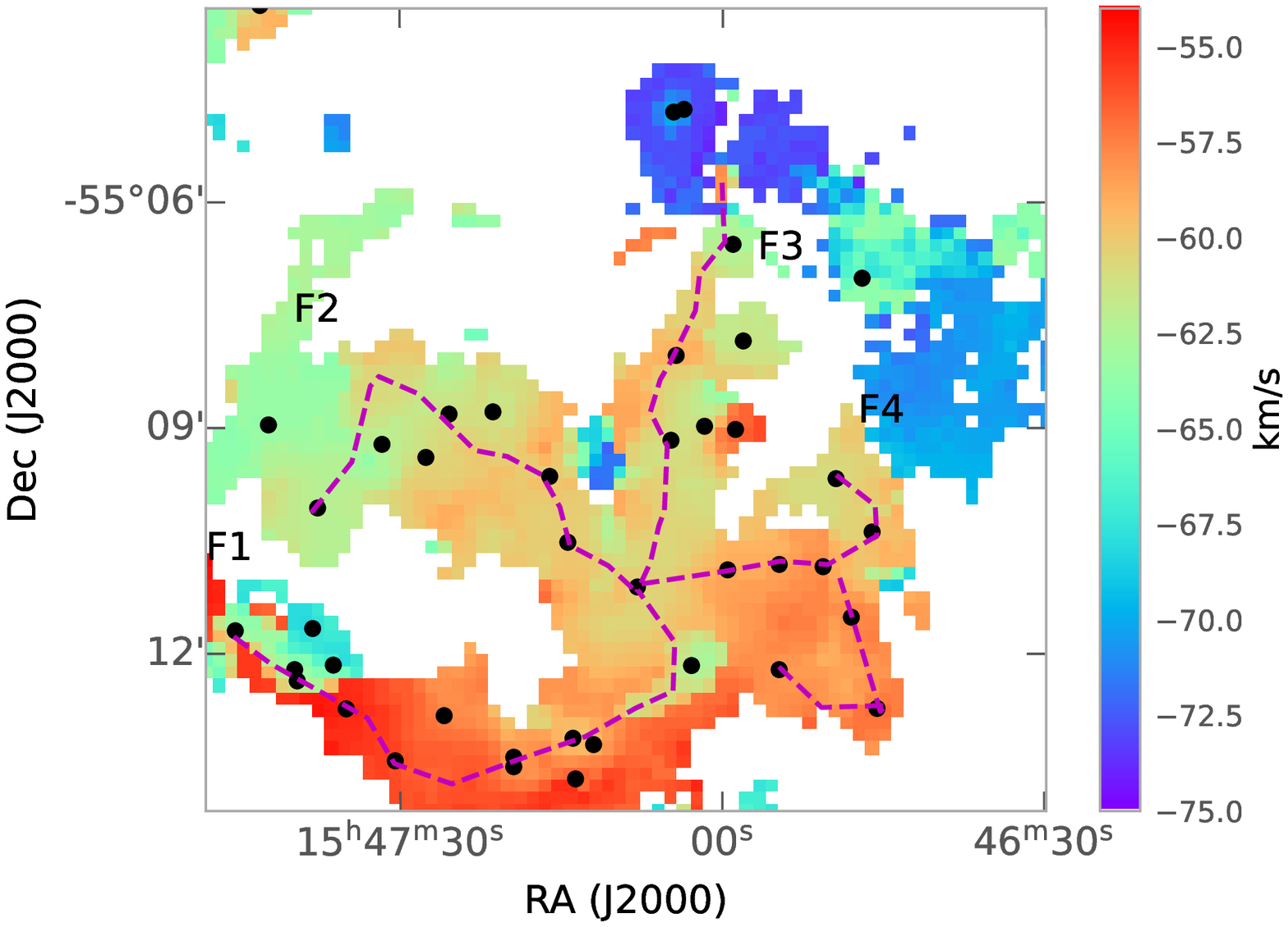}
\includegraphics[width=0.52\textwidth]{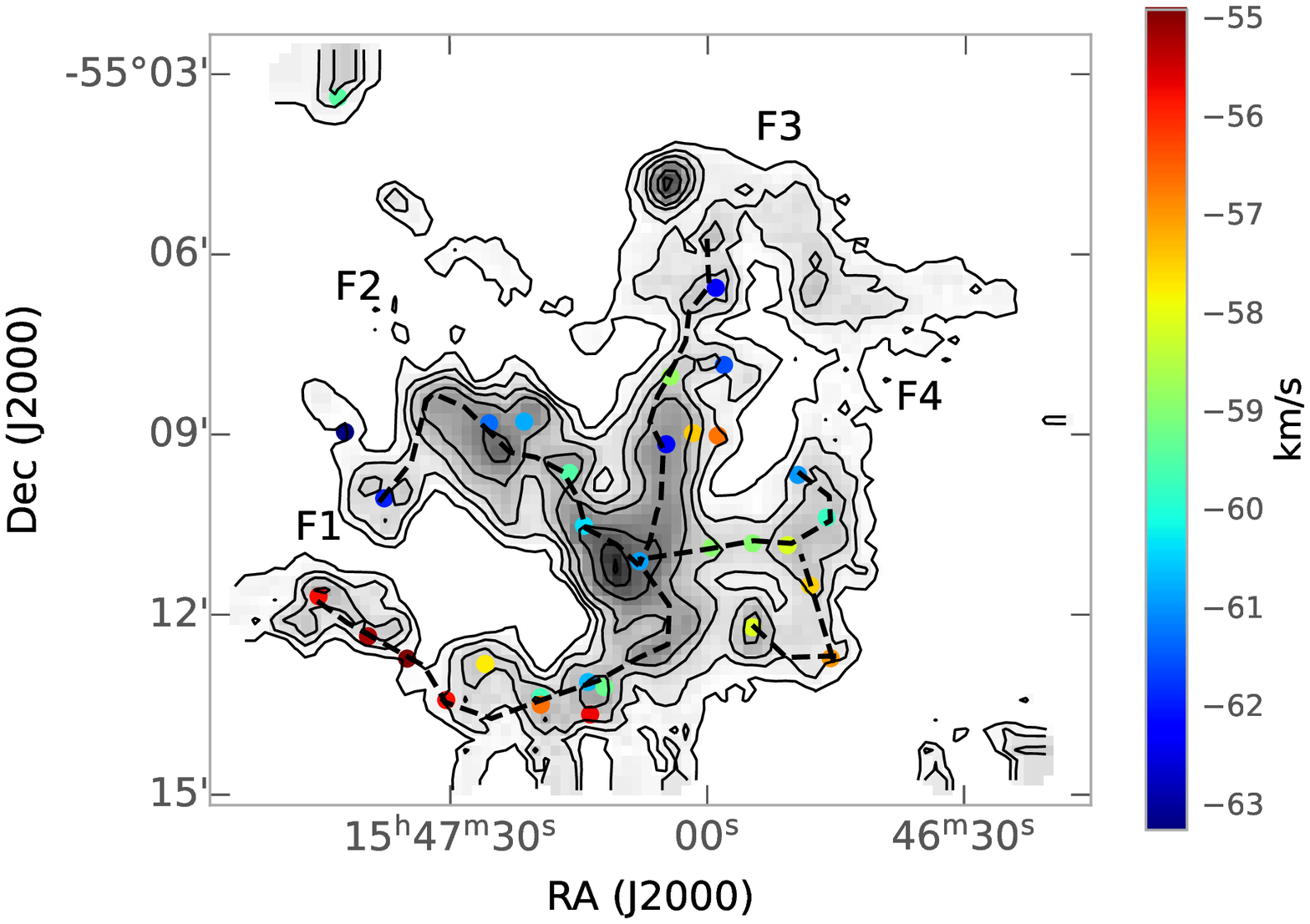}
\caption{(Left) Velocity distribution (moment 1) of the region from
CO(3--2) data shown along with the filaments identified in the region.
(Right) Integrated intensity (moment zero) map of the G326 region overlaid with the four filaments identified
in the region. The distribution of $^{13}$CO clumps extracted using the dendrogram analysis is also shown.
Only clumps with velocities above -63\,\kms are shown in the figure. The clumps are colour coded according to their velocities according to the colorscale shown to the right.
\label{fig_spiral}}
\end{figure*}

The large scale velocity distribution of the HFS and the sub-structures as obtained from the moment 1 map of CO(3--2)  shows a morphology with four main arms forming a spiral structure consistent with the velocity coherent filaments identified in the channel maps using the filfinder method (Fig.\,\ref{fig_spiral} (Left)). The northern part of the HFS is relatively blue- shifted with respect to the central region whereas towards the south, velocities are more red-shifted. The velocities across the different filaments fall within $\sim\pm$ 5\,\kms of the central hub except towards the north-west, where the emission from a secondary molecular cloud is identified with a blue-shifted velocity of up to 15\,\kms with respect to the main cloud. The $p$-$\upsilon$ diagrams discussed in Sec. 5.3 show signatures of interaction of this secondary cloud with the cloud hosting the HFS in G326. Hence, we consider the HFS and the secondary cloud together while discussing the large scale morphology of the region. 

The thirty-nine clumps identified using the dendrogram analysis are found to be
concentrated along the four arms of the cloud, identified as the four
filaments F1, F2, F3 and F4 numbered in the clockwise direction
(Fig.\,\ref{fig_spiral} (left)). To study the velocity structure of the individual filaments belonging to the primary cloud, we have excluded all the clumps belonging to the secondary cloud (i.e., V< -65\,\kms) from further analysis. The distribution (Fig.\,\ref{fig_spiral} (Right)) follows a trend that all the clumps coinciding with the southern filaments (F1 and F4) are relatively red-shifted with respect to the clump in the central hub whereas the northern clumps in filaments F2 and F3 are blue-shifted, which is consistent with the moment 1 map.

The spiral morphology observed here can arise from cloud rotation or convergence of filaments towards the central hub. Similar spiral-like morphology is also observed in many Galactic clouds associated with high mass star forming regions \citep[e.g.,][]{{2016ApJ...828...32L},{2017A&A...598A..96L},{2019A&A...628A...6S},{Trevino2019}} on similar spatial scales. This indicates that spiral structures  are common in molecular clouds and are possibly more ubiquitous than previously thought. \citet{Wang2022} have discussed the possible origin of such spiral structure based on the pattern of magnetic fields in the regions as well as on signatures of cloud-cloud collisions. For the HFS associated with G326 in the absence of polarimetric data and any clear indication of a "collision" as opposed to the interaction (Sec. 5.3) that we find, it is not possible to obtain better clarity on the origin of the spiral structure in the region.  We further emphasize that systematic statistical study of such spiral features in a much bigger sample will be necessary to understand their origin.

\begin{table}
\begin{center}
\caption{Mass accretion rates for filaments in G326 \label{tab_accrate}}
\begin{tabular}{cccc}
\hline
\hline
Filament& Mass &V$_\mathrm{grad}$ & $\dot{M}_\mathrm{fil}$\\
 & M$_\odot$ &\kms\,pc$^{-1}$ & M$_\odot$\,yr$^{-1}$\\
\hline
F1 & 589 &0.6&$3.6\times10^{-4}$\\
F2 & 757 &0.4&$3.1\times10^{-4}$\\
F3 & 287 &0.6&$1.8\times10^{-4}$\\
F4 & 305 &0.4&$1.2\times10^{-4}$\\
\hline
\end{tabular}
\end{center}
\end{table}

We estimate the mass accretion rate along the filaments using the mass of the individual filaments and the estimated velocity gradients from the position-velocity diagrams (Fig.\,\ref{fig_pvdiag}, \ref{fig_velprof}) along the filaments (details in Sec. 5.2). The mass of the individual filaments are estimated from $^{13}$CO(3--2) emission map assuming the average $T_{\rm ex} = 25$\,K and $\tau_{13}$=0.5 leading to a CO to H$_2$ conversion factor of 4.4$\times 10^{20}$\,\cmsq / K\,\kms, as discussed in Section 4.2. Assuming a simple cylindrical model for the filaments in G326, the mass accretion rate are calculated using the expression given by \citet{2013ApJ...766..115K}

\begin{equation}
\dot{M}_{\rm fil} = \frac{V_{\rm grad} M_\mathrm{fil}}{\tan\alpha}    
\end{equation}

\noindent 
where $V_{\rm grad}$ is the velocity gradient along the filament as measured from the $p$-$\upsilon$ diagrams, $M_{\rm fil}$ is the mass of the filament and $\alpha$ is the angle of inclination of the long axis of the cylinder with respect to the plane of the sky. We have considered an inclination angle of 45\arcdeg\ for our calculation. Assumption of $\alpha$ as 25\arcdeg\ increases the accretion rate by a factor of 2 whereas a value of  65\arcdeg\ for $\alpha$ decreases the accretion rate to half. The accretion rates of all the filaments in G326 HFS are listed in Table~\ref{tab_accrate} and ranges from 1.2--3.6$\times10^{-4}$~M$_\odot$\,yr$^{-1}$. The values are consistent with typical accretion flow rates found in other Galactic HFS \citep[$10^{-4}-10^{-3}$~M$_\odot$\,yr$^{-1}$ ;][]{Trevino2019,Chen2019,Liu2021}.

\section{Discussion: Outflows and Star formation activity in S6, S7 and S10}

\begin{figure*}
\includegraphics[width=0.32\textwidth]{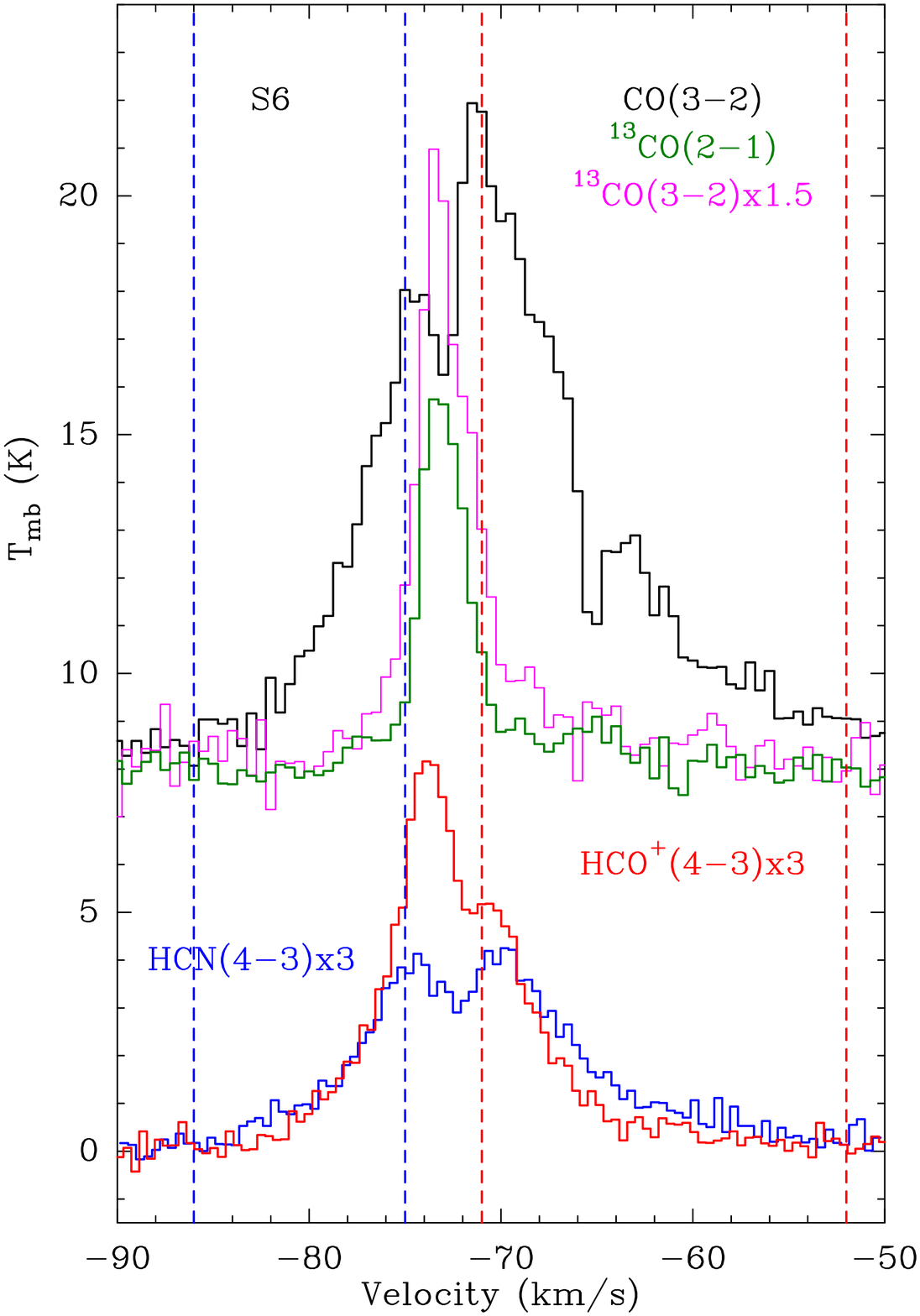}
\includegraphics[width=0.30\textwidth]{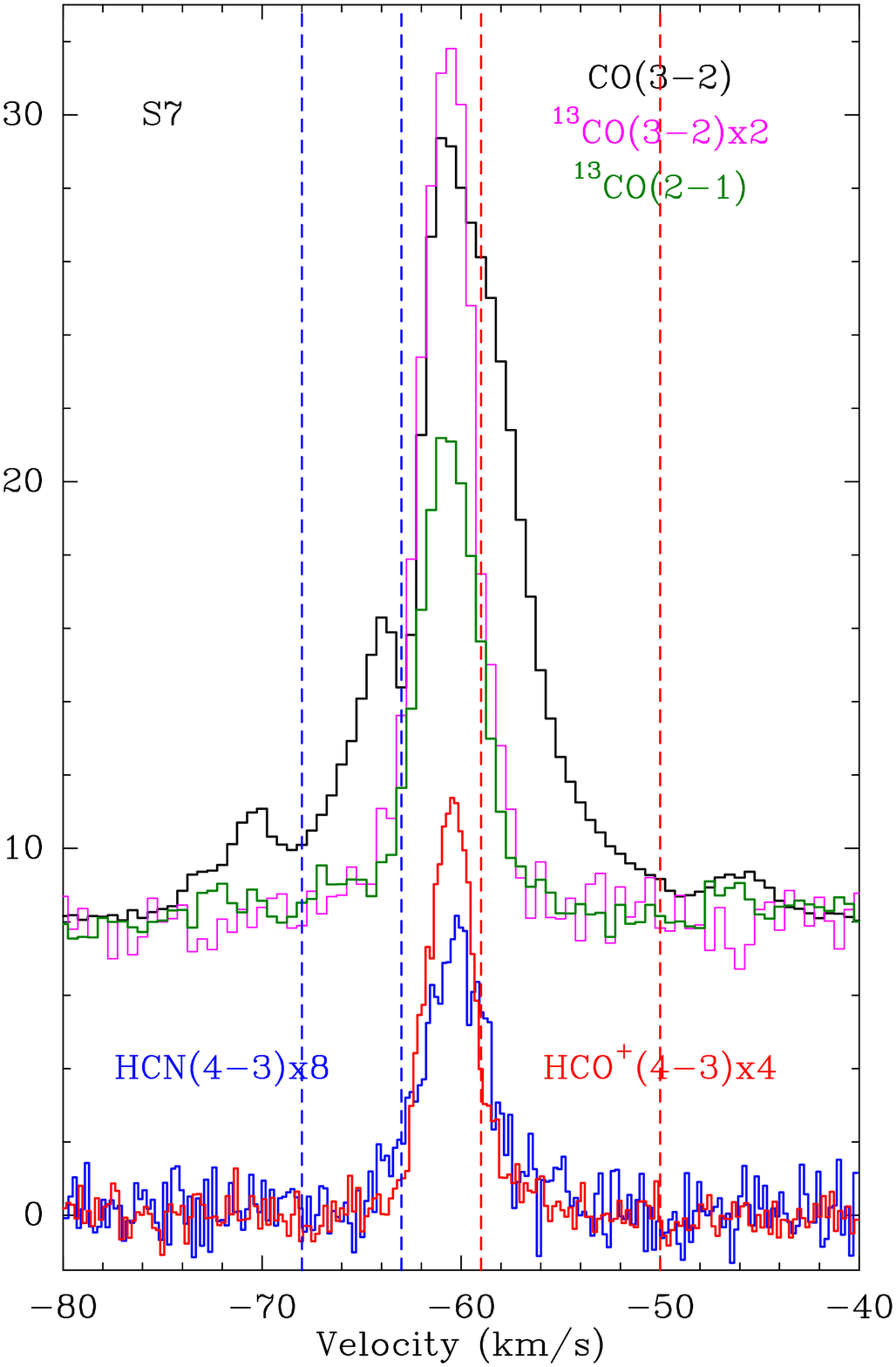}
\includegraphics[width=0.30\textwidth]{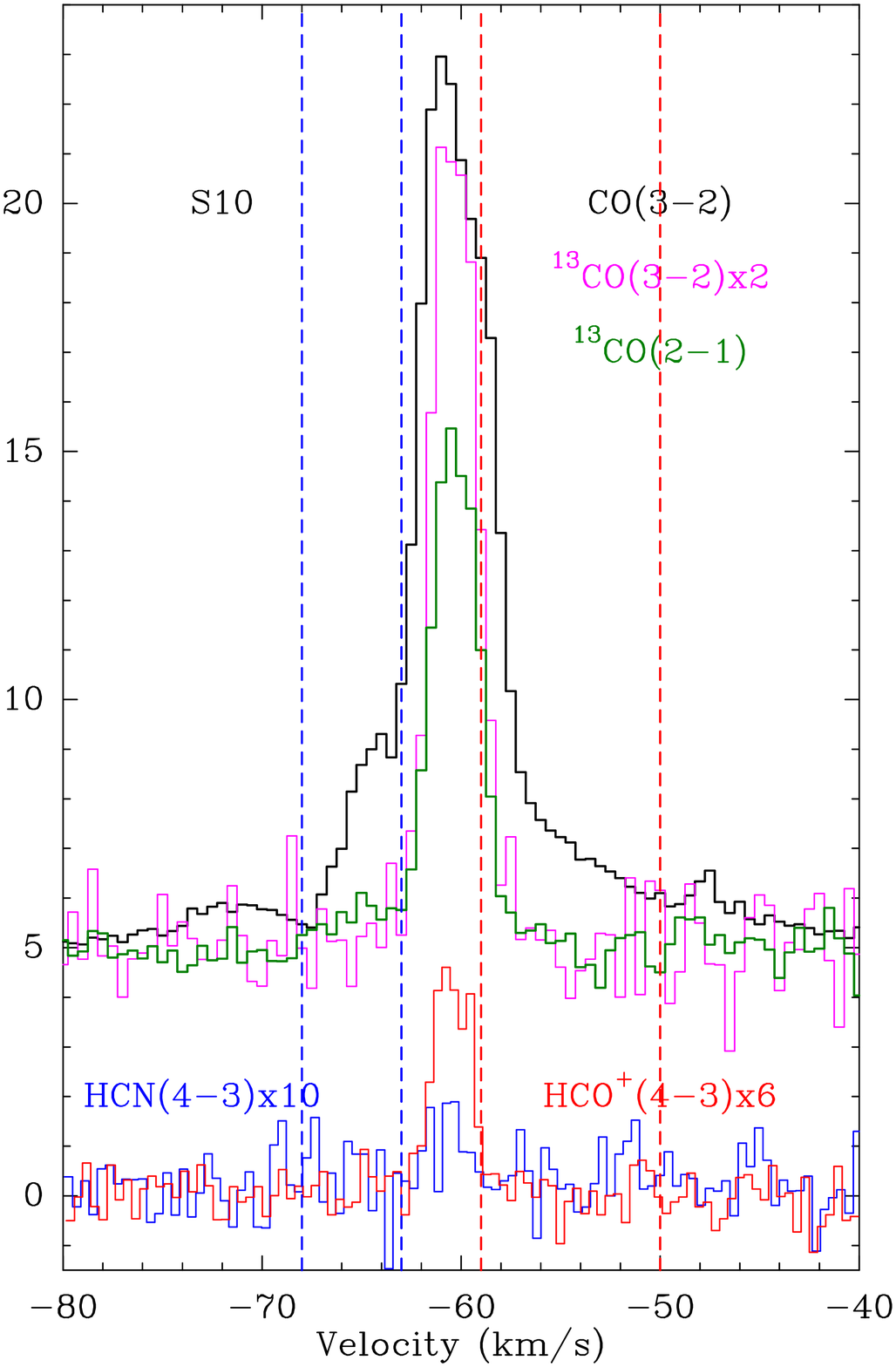}
\caption{Comparison of spectra at the peak positions of the protostellar 
sources \#6,\#7 and \#10. Spectra have been multiplied and shifted by appropriate values for better visibility as marked in the panels.
\label{fig_ysospec}}
\end{figure*}

\begin{figure}
\centering
\includegraphics[width=0.50\textwidth]{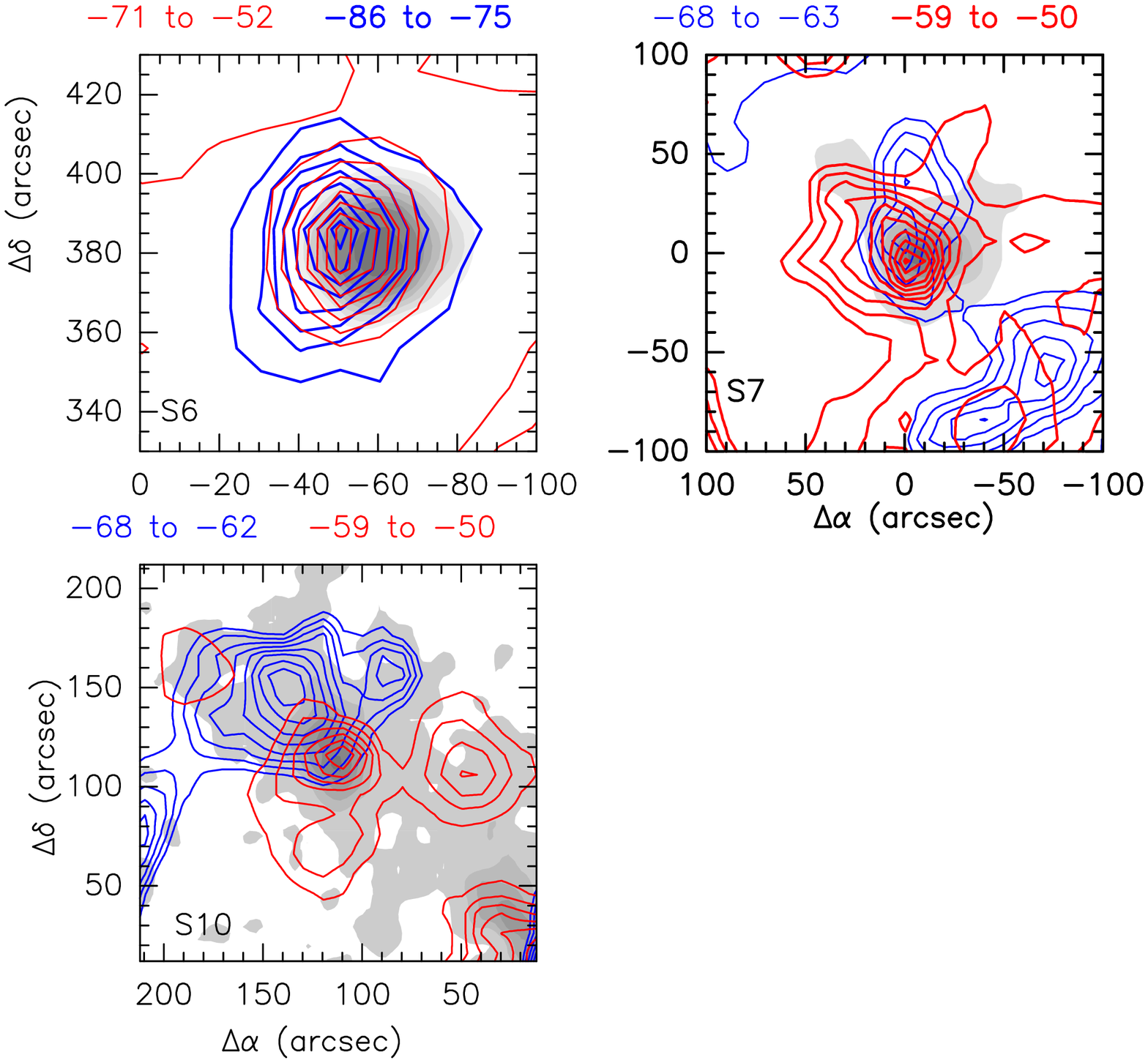}
\caption{Maps of gas entrained in the outflows corresponding to the
sources S6, S7 and S10. The red and blue contours show the intensity
distribution of the two lobes of the outflows with values written as
minimum to maximum (stepsize).  Blue contour levels (in K\,\kms) are for
S6 (15 to 43 (5)), S7 (10 to 30 (5)), S10 (20 to 40 (5)). Red contour
levels (in K\,\kms) are for S6 (15 to 90 (5)), S7 (15 to 60 (5)), S10
(10 to 35 (5)).
\label{fig_outfmap}}
\end{figure}

We study the properties of three most massive clumps identified in the
region, CL11, CL15 and CL22 (Table\,\ref{tab_dendro}), of which two are
in the southern filament with velocities around -61\,\kms\ and one is in
the northern filament with $\upsilon_{\rm LSR}\sim -73$\,\kms.  The
clumps CL11, CL15 and CL22 are associated with the Hi-GAL sources S6,
S7, and S10 (marked in Fig\,\ref{fig_contover}) respectively.
\citet{Urquhart2022} has classified the sources S7 and S10 as YSOs and
S6 as a protostellar object. The position-velocity diagram along the cut
AB (Fig.\ref{fig_pvdiag}) shows that at the location of the protostellar
sources S6, S7 and S10 (Table\,\ref{tab_fir}), there is significant
broadening of the spectra strongly indicative of outflow activity. We
compare the spectra of CO(3--2), \thCO(3--2), \thCO(2--1),
HCO$^+$(4--3),  HCN(4--3) at the position of the three protostellar
sources (Fig.\,\ref{fig_ysospec}). For all sources the CO(3--2) spectra
clearly show the broadening due to the outflow. For the source S6, the
outflow appears to be the most pronounced and clearly seen also in the
HCN(4--3) and HCO$^+$(4--3) spectra, with the CO(3--2) spectrum showing
a clear self-absorption dip. Identification of the red- and blue-wings
of the outflows in the CO(3--2) spectra is complicated by the presence
of the additional velocity components due to the diffuse gas.
Figure\,\ref{fig_outfmap} shows maps of the red- and blue lobes of the
outflows around the protostellar sources, the velocity ranges of
integration for the lobes are marked with red and blue dashed vertical
lines in Fig.\,\ref{fig_ysospec}. The blue and red lobes of the outflow
for sources S6 and S7 appear to be centered on the peak of the continuum
emission and the nearly circular contours of the source S6 suggest that
the axis of the outflow is nearly aligned with the line of sight of the
observer. As is evident from Fig.\,\ref{fig_outfmap} the blue lobes of the
outflows for both S6 and S7 are fainter than the red lobe, for S10 the two 
lobes have similar peak intensities but do not overlap spatially.

We have determined the properties of the outflows associated with the sources, S6, S7 and S10 using the velocities and extents of the blue- and red-shifted lobes of the outflows (Figs\,\ref{fig_pvdiag}, \ref{fig_outfmap}). Based on the integrated intensity maps of the outflows (Fig.\,\ref{fig_outfmap}), we assume the outflows associated with S7 and S10 to be inclined at an angle of 57\arcdeg, while for  S6, which appears to be certainly directed along the line-of-sight, the angle is assumed to be 10\arcdeg. The total CO column density of each pixel in the outflows was estimated  assuming $T_{\rm ex}$ = 20\,K, and the outflow wings to be optically thin using:

\begin{equation}
{\rm N_{tot}(^{12}CO)} = \frac{3k^2_{\rm B}T_{\rm ex}}{4\pi^3 \mu_{d}^2 h\nu^2 \exp(-2h\nu/k_{\rm B}T_{\rm ex})}\displaystyle\int {T_{\rm mb}{\rm d}\upsilon}
\end{equation}

where $k_{\rm B}$=1.38$\times 10^{-16}$\,erg\,K$^{-1}$, $h$=6.626$\times 10^{-27}$\,erg, $\mu_{\rm d}$ = 0.112$\times 10^{-18}$\,esu\,cm, $\nu$=345.79599\,GHz, and $\upsilon$ is in \kms.

We derive the total mass ($m_{\rm out}$) of each lobe by integrating up to the lowest contour level as indicated in Fig.\,\ref{fig_outfmap}, 
The mass for each pixel in the deﬁned outﬂow lobe area is computed using
\begin{equation}
M_{\rm pixel} = N_{\rm tot}(^{12}{\rm CO}[{\rm H_2/CO}]\mu_{\rm H_2}m_{\rm H} A_{\rm pixel}    
\end{equation}

where $m_{H_2}$ = 2.72 is the mean molecular weight, $m_{\rm H}$ = 1.67$\times 10^{-24}$\,g is the mass of a hydrogen atom, [CO/H$_2$] is assumed to be 10$^{-4}$, and $A_{\rm pixel}$ is the area of each pixel within the outﬂow lobe so deﬁned.

Since the emission from the region is contaminated by the presence of a
blue- and a red-shifted velocity component it is difficult to clearly
assign a maximum velocity to the lobes of the outflow. For all
calculations we thus use the central velocity in the interval over which
a lobe of an outflow is detected as the mean velocity
($\upsilon_b$ or $\upsilon_r$) of that lobe.  For the estimate of the
dynamic timescale ($t_{\rm dyn}$) of outflows we have estimated the size
of the outflow from the beam deconvolved size of the lowest
contour drawn in Fig.\,\ref{fig_outfmap}. Using the range of velocities
($\upsilon_{\rm out}$)
over which the lobe of the outflow is seen, we estimate the momentum,
energy and luminosity of the outflow as follows:

\begin{align}
P_{\rm out} & =  m_{\rm out} \times \upsilon_{\rm out}\\
E_{\rm out} & =  \frac{1}{2} m_{\rm out} \times \upsilon_{\rm out}^2\\
L_{\rm out} & =  E_{\rm out}/t_{\rm dyn}
\end{align}

Table\,\ref{tab_outflow} presents the size, the dynamical timescale and
the total mass, momentum, energy and luminosity of the red and blue
lobes of the outflows associated with the sources S6, S7, and S10.

\begin{table*}
\begin{center}
\caption{Properties of outflows from CO(3--2) in G326 \label{tab_outflow}}
\begin{tabular}{ccccccccccccc}
\hline
\hline
Source & 
\multicolumn{2}{c}{Size} & 
\multicolumn{2}{c}{$t_{\rm dyn}$} & 
\multicolumn{2}{c}{Mass} & 
\multicolumn{2}{c}{Momentum} & 
\multicolumn{2}{c}{Energy} & 
\multicolumn{2}{c}{Luminosity} \\
& Blue & Red & Blue  & Red & Blue  & Red & Blue & Red & Blue & Red & Blue & Red\\
&(\arcsec) & (\arcsec) & yr & yr & \msun & \msun & \msun \kms & \msun \kms & erg & erg & \lsun &
\lsun \\
\hline
S6 &  32 & 24 & 8.3$\times 10^4$ & 3.7$\times 10^4$ & 1.9 & 3.6 & 14.7 & 45.6  &
1.1$\times$10$^{45}$ & 5.7$\times 10^{45}$ & 0.11 & 1.20 \\
S7 & 32 & 44 & 1.1$\times 10^5$ & 1.3$\times 10^5$ & 2.2 & 4.5 & 20.6 & 49.3 &
1.9$\times 10^{45}$ & 5.4$\times 10^{45}$ & 0.14 & 0.35\\
S10 & 34 & 41 & 1.3$\times 10^5$ & 1.3$\times 10^5$ &  3.2 & 2.0 & 26.4 & 20.8 &
2.2$\times 10^{45}$ & 2.1$\times 10^{45}$ & 0.13 & 0.14\\
\hline
\end{tabular}
\end{center}
\end{table*}

Using the {\em VIZIER} photometry viewer and Table\,\ref{tab_fir} we have compiled the available
photometric measurements of S6, S7 and S10 at $\lambda >3$\,\micron. Numerical integration of
the spectral energy distribution (SED) thus obtained gives luminosities of  2320, 2920 and 260\,\lsun\ for S6, S7 and S10, respectively. Comparison of the luminosities thus estimated, with the luminosities 
calculated for Zero Age Main Sequence (ZAMS) stars \citep{thompson1984}, suggest that the stellar types of S6, S7 and S10 are B2, B2.5 and B6 respectively. This method could overestimate the spectral type 
since dust in the region may be heated by lower mass stars in the cluster in addition to the massive YSOs
ionizing UC \HII\ regions.  The luminosities of the sources are adequate to provide the moderate luminosities observed in the outflows. 

The masses of the molecular clumps CL11 (S6), CL15 (S7) and CL22 (S10)
as derived from the dendrogram analysis are 413, 187 and 535\,\msun\
respectively. The average gas density ($\rho$) for the clumps S6, S7
and S10 derived from these mass and sizes are 1.1$\times 10^5$,
6.6$\times 10^4$, and 2.7$\times 10^4$\,\cmcub, respectively. We have
compared the velocity-integrated peak intensities of HCO$^+$(4--3) and
HCN(4--3) at the positions S6 and S7 with the results of the non-LTE
radiative transfer code RADEX \citep{VdTak2007} to constrain the gas
densities at these positions. We estimate $I$(HCO$^+$(4--3)) to be
20.2 and 8.7\,\Kkms\ at S6 and S7 respectively, and $I$(HCN(4--3)) to
be 17.4 and 4.1\,\Kkms\ for the same sources in the same order. Radex
calculates line intensities as a function of density, kinetic
temperature and column density of the species. Based on the available
literature and this work, both sources S6 and S7 are not detected in
the radio continuum, but are bright in the far-infrared continuum.
Thus, if we consider the four broadly identified stages of massive
star classification, {\em viz.,} Infrared Dark Clouds (IRDC), High
Mass Protostellar Objects (HMPO), Ultracompact \HII\ regions (UCHII)
and Hot Molecular Core (HMC), then sources S6 and S7 with no radio
emission detected so far are likely to be HMPOs. For comparison with
the results of radex, we adopt for S6 and S7 the column densities of
HCN and HCO$^+$ to be the same as the median column densities of
9.1$\times 10^{13}$\,\cmsq\ and 1.2$\times 10^{14}$\,\cmsq\
respectively as observed in HMPOs by \citet{Gerner2014}. Since the two
transitions have almost identical upper energy levels, the intensity
ratios for given column densities is not sensitive to $T_{\rm kin}$
and the RADEX analysis indicates peak densities of  4$\times
10^6$\,\cmcub\ and 10$^6$\,\cmcub\ for S6 and S7 respectively. For the
source S10 where only HCO$^+$(4--3) is detected, based on the critical
densities of 8.5$\times 10^6$\,\cmcub\ and 1.8$\times 10^6$\,\cmcub\
of HCN(4--3) and HCO$^+$(4--3) we estimate the densities to be
$<10^6$\,\cmcub.

From the average densities derived from the clump masses (Table\,\ref{tab_dendro})  we estimate the free-fall times ($\sqrt{\frac{3\pi}{32G\rho}}$) of the cores S6, S7 and S10 to be 1.1$\times 10^5$, 1.4$\times 10^5$ and 2.2$\times 10^5$ yr
respectively. The free-fall times are an estimate of the timescale for
dynamical evolution of the core and theoretical studies of  core
evolution suggest that the timescale for star formation in a core is
of order of a few free-fall times (e.g., Tan \& McKee 2002). For the
three sources the dynamical timescales of the outflow and the
free-fall timescales of the core appear to be of similar magnitudes.
The free-fall velocities estimated for the clumps S6, S7 and S10 using $\sqrt{\frac{2GM_{\rm cl}}{Radius}}$ are 3.7, 2.6 and 3.2\,\kms\ respectively. These velocities are 
comparable to the linewidths, ($\Delta \upsilon = 2.354\sigma_{\upsilon}$) 
determined from the dendrogram analysis. 
We estimate the gravitational potential energies ($E_{\rm grav}$) of
the cores using $G M^2_{\rm core}/R_{\rm core}$ for the masses and
sizes determined from the dendrogram analysis. The $|E_{\rm grav}|$
for S6, S7 and S10 are estimated to be 1.9$\times$10$^{46}$\,erg,
1.1$\times$10$^{46}$\,erg and 4.9$\times$10$^{46}$\,erg respectively.
The kinetic energy of the outflows ranging between 4--7.5$\times
10^{45}$\,erg (Table\,\ref{tab_outflow}) is clearly smaller than
$E_{\rm grav}$, thus suggesting that the collapse of all three regions
would continue unless there is additional support from magnetic field 
or turbulence.  We note that at a resolution of 0.3--0.4\,pc it is
possible that the region identified as a single core corresponds to a
clump (as opposed to a core), so a direct comparison between the
outflow energies with gravitational energies should only be regarded
as an estimate. The virial parameters of the sources S6, S7 and S10 lie between 1.6--2.7 and are generally smaller than the values found for the other clumps in the region. While these values for virial parameters for the clumps indicate support against gravitational collapse, the association with star-forming far-infrared sources implies the presence of collapsing cores inside the clumps. Finally, based on the masses of the clumps (Table\,\ref{tab_dendro}) and their bolometric luminosity (Table\,\ref{tab_fir}) we identify MYSOs that are forming at the hub as well as at the end of the filament F2 (S10).

\section{Summary}

We have studied the massive star forming environment and the gas dynamics state of the Hub-Filament-System associated with G326.27-0.49. The newly obtained $J$=3--2 CO(\thCO) observations clearly show (i) clouds emitting at two distinct velocities of -61\,\kms\ and -72\,\kms\ and the presence of gas at intermediate velocities as seen in the $p$--$\upsilon$ diagrams indicate interaction between the two parts and (ii) the -61\,\kms\ cloud consisting of filaments that can be grouped into four main arms  through which gas accretion happens in the hub harboring a protostellar object S7 with a strong outflow. The filaments, typically show velocity gradients of around 0.5\,\kms\,pc$^{-1}$, and take up the form of a spiral structure as has recently been seen in several other such regions. Among these filaments, the filaments F2, on which the YSOs S7 and S10 lie is the most massive and also registers the second largest mass accretion rate.  Based on a  dendrogram-based analysis we identify a total of 39 and 25 clumps in the CO(3--2) and \thCO(3--2) maps, of which most of the optically thin \thCO(3--2) clumps appear to have large virial parameter not indicative of susceptibility to gravitational collapse. We detect three sources, S6, S7 and S10 with strong outflows in CO(3--2) and suggest that the estimated $\alpha_{\rm vir}$ of 1.6--2.7 for these sources is likely a result of the fact that at the resolution (0.4\,pc) of our observations we are observing the clumps rather than gravitationally contracting cores that give rise to the outflows. Comparison of the energy in the outflows and the gravitational energy of the embedded sources and also the linewidth of clumps with free-fall velocities suggest that the sources are in a state of collapse.  High density tracers such as $J$=4--3 transitions of HCO$^+$ and HCN suggest densities in excess of 10$^6$\,\cmcub\ of the YSOs S6 and S7. The source S6, in the northern part of the region, has a luminosity consistent with a B2 star and drives the strongest outflow in the region with a total luminosity of $1.3$\,\lsun. The YSO S7 currently forming at the hub of the HFS has the luminosity of a B2.5 star, and drives an outflow with a total luminosity of 0.49\,\lsun. Our study of the HFS associated with G326.27-0.49 suggests flow of gas along the filaments towards the massive star forming hub, as well as presence of young high-mass protostars at the ends of filaments F2 and F3. This is consistent with the findings of recent studies of high-mass star forming regions which favour accretion from a larger mass reservoir \citep[][and references therein]{Hacar2022}, thus providing support to the clump-fed as well as edge collapse of filaments as possible scenarios of high-mass star formation. The moderate resolutions of the observations presented here enable us to obtain an overall perspective of the area around G326.27-0.49 as a region with multiple filaments, four of which intersect at the hub, where massive star formation is in progress. More detailed characterization to constrain the flow of matter and evolution of the filaments as well as the cores, require observations that spatially resolve the star forming cores as well as the filaments.

\section*{Acknowledgements}
BM acknowledges the support of the Department of Atomic Energy, Government of India, under Project Identification No. RTI 4002. This publication is based on data acquired with the Atacama Pathfinder Experiment (APEX) under programmes 092.F-9315 and 193.C-0584. APEX is a collaboration among the Max-Planck-Institut fur Radioastronomie, the European Southern Observatory, and the Onsala Space Observatory. The processed data products are available from the SEDIGISM survey database located at https://sedigism.mpifr-bonn.mpg.de/index.html, which was constructed by James Urquhart and hosted by the Max Planck Institute for Radio Astronomy. This research has made use of the VizieR catalogue access tool, CDS, Strasbourg, France (DOI : 10.26093/cds/vizier). The original description of the VizieR service was published in 2000, A\&AS 143, 23"



\section*{Data Availability}

This paper uses archival data that are publicly available. The new data observed with APEX that is presented here will be shared on reasonable request to the corresponding author.

\bibliographystyle{mnras}
\bibliography{g326} 

\appendix

\section{Dust Temperature Map of G326}

\begin{figure}

\includegraphics[width=0.48\textwidth]{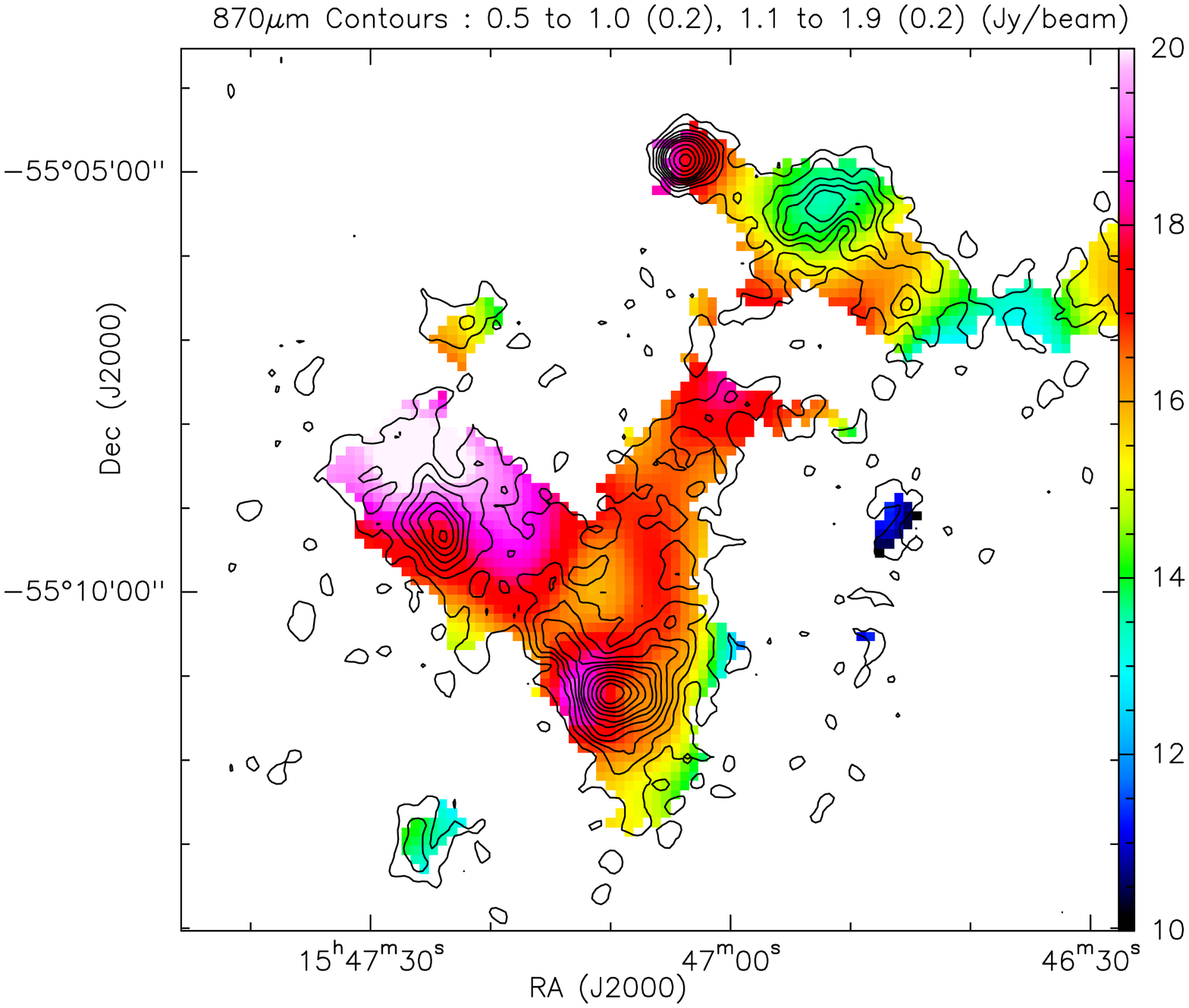}
\caption{Map of dust temperature in the region obtained from
pixel-by-pixel fitting of 160, 250, 350 and
870\,\micron\ continuum emission with a grey-body function for a dust
emissivity exponent ($\beta$) of 2.0 shown in color. The Planck-ATLASGAL
map of 870\,\micron\ continuum emission is plotted as contours with
levels mentioned above the panel.
\label{fig_dtemp}}
\end{figure}

\section{Identification of filaments using Filfinder}

\begin{figure*}
\includegraphics[width=0.95\textwidth]{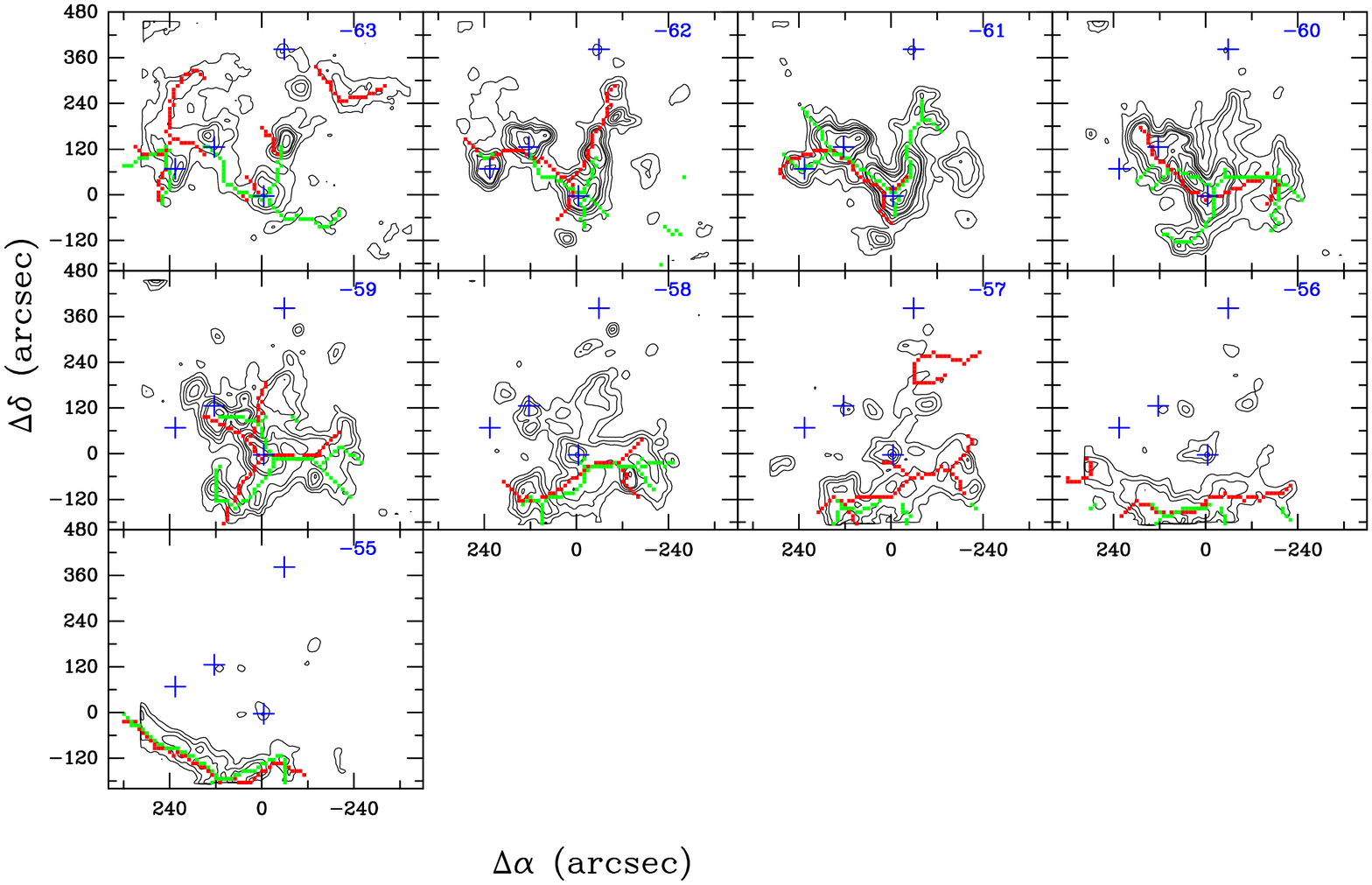}
\caption{Filaments identified in the different velocity slices of the CO(3--2) (in green) and \thCO(3--2) (in red) data shown on the CO(3--2) channel map in the velocity interval -63 to -55\,\kms. The contours levels are 2, 4, 6, 8.5, 10, 14, 18, 22 and 24\,K. Channel velocities indicated on the top right corner and the blue '+' signs show the location
of S6, S7, S10 and S15. Rest of the details same as in Fig.\,\ref{fig_chanmap_co32}
\label{fig_filfinder}}
\end{figure*}

The filamentary velocity-coherent structures in the region were identified by applying the python package FilFinder \citep{Koch2015} on both the CO and \thCO(3--2) datacubes at a resolution of 1\,\kms. The FilFinder algorithm segments filamentary structure by using adaptive thresholding, which performs thresholding over local neighborhoods and allows for the extraction of structure over a large dynamic range. Input parameters for FilFinder include: (1) global threshold -- the minimum intensity for a pixel to be included; (2) adaptive threshold -- the expected full width of filaments for adaptive thresholding; 
(3) smooth size -- scale size for removing small noise variations; (4) size threshold -- minimum number of pixels for a region to be considered as a filament. The emission structures in each velocity bin were first flattened to 95 percentile to smooth the bright features in the image.  While creating masks, the global threshold was set at the 45th percentile for that velocity, the adaptive threshold was set at 6.9\,pc in order to capture the structures seen in the channel maps by eye, the size threshold was set at 2000 arcsec$^2$ and the smooth size was set to 0.4\,pc. For comparison, at the distance to the source the spatial resolution of the CO(3--2) map is 0.40\,pc. Figure\,\ref{fig_filfinder} shows the medial axes of the filaments identified in the different velocity planes of the CO(3--2) and \thCO(3--2) datacubes. 

\section{CO Excitation Temperature and $^{13}$CO optical depth map}

\begin{figure*}
\includegraphics[width=0.90\textwidth]{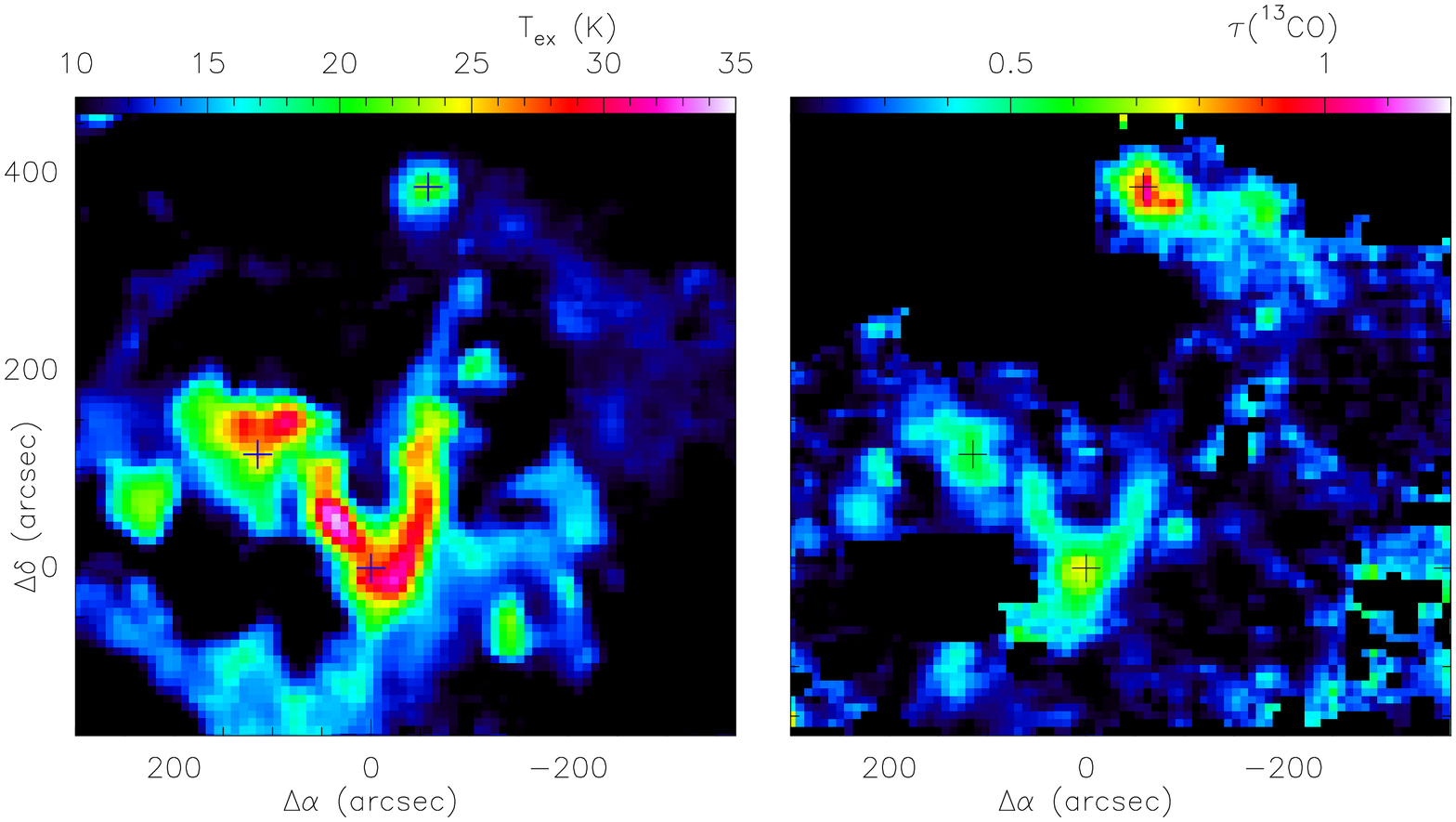}
\caption{Maps of ({\em Left}) excitation temperature derived from $T_{\rm mb}$(\twCO(3--2)) emission assuming the the CO emission to be optically thick and LTE (Eq. 4) and ({\em Right}) Optical depth of \thCO, $\tau_{13}$ estimated from the excitation temperature distribution and $T_{\rm mb}$(\thCO(3--2)) using Eq. 6. The
coordinates are shown as offsets in arc seconds relative to the center
RA:15$^{\rm h }$47$^{\rm m}$10.8$^{\rm s}$ Dec:-55\arcdeg 11\arcmin 12\arcsec. The '+' in both plots show the positions of the sources S6, S7 and S10.
\label{fig_tex}}
\end{figure*}

We have used the \twCO(3--2) and \thCO(3--2) maps to estimate the excitation temperature ($T_{\rm ex}$) and optical depth ($\tau_{13}$) distribution in the region (Fig.\,\ref{fig_tex}).

\section{Detailed results of dendrogram analysis \label{sec_appdendro}}

\begin{figure*}
\includegraphics[width=0.48\textwidth]{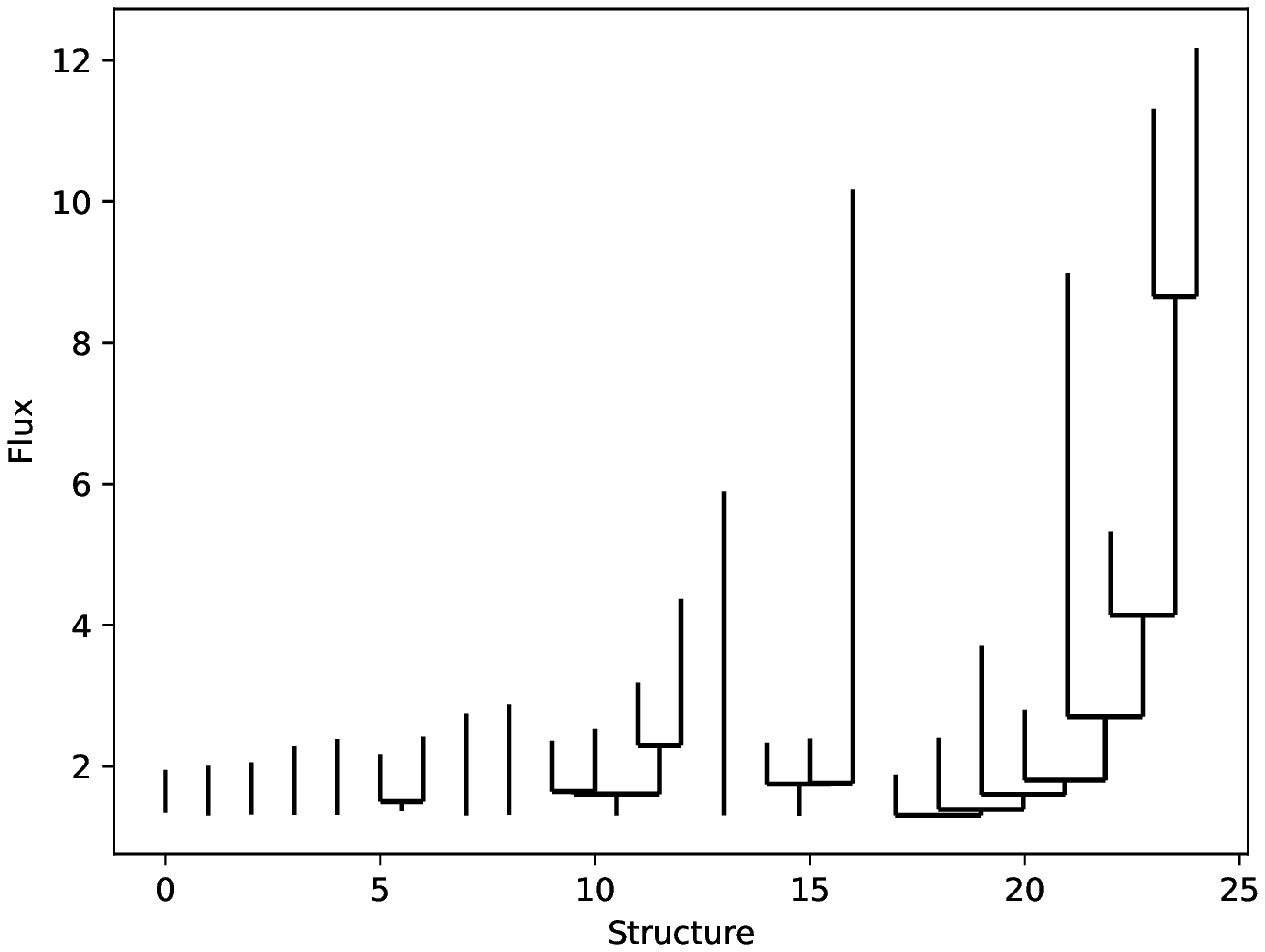}
\includegraphics[width=0.48\textwidth]{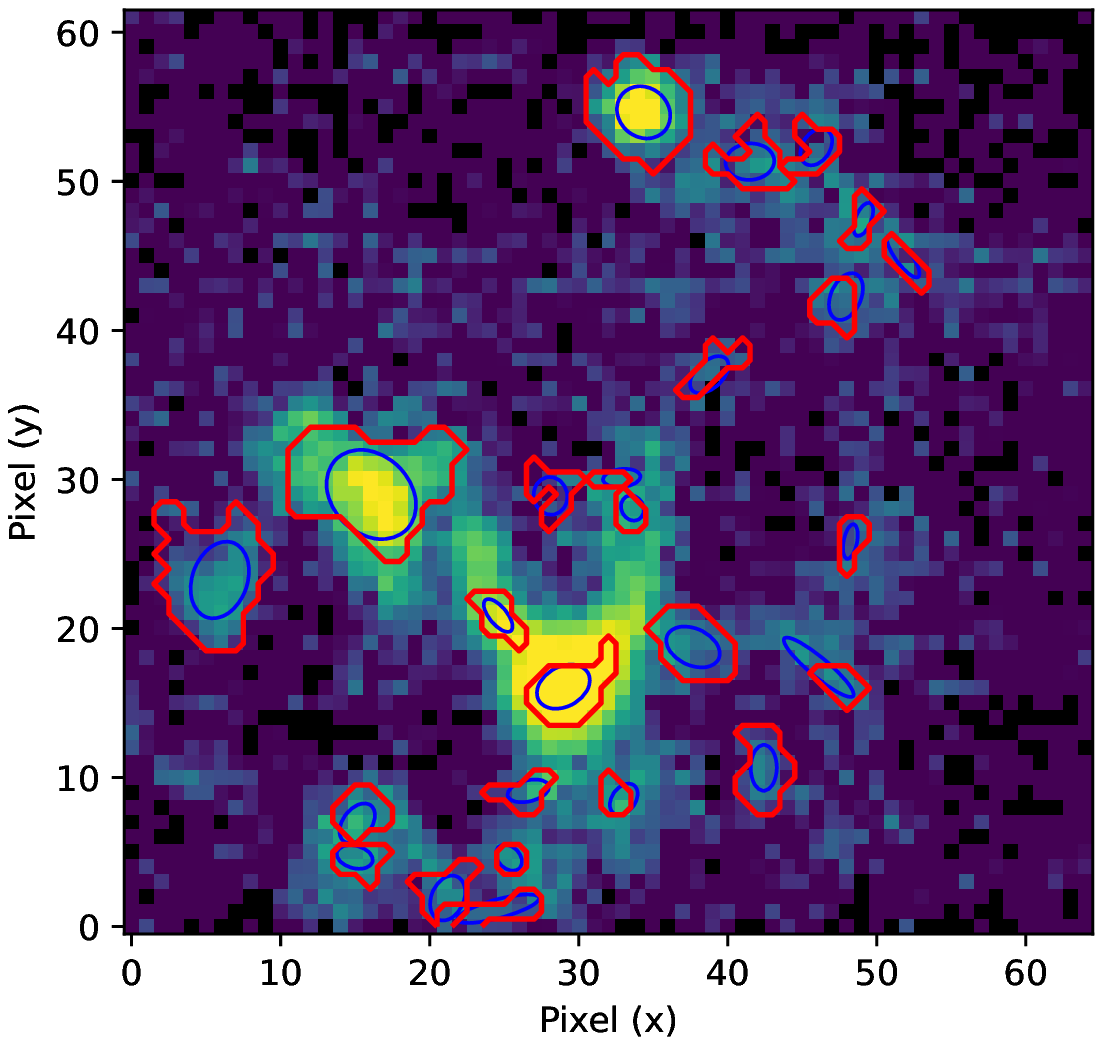}
\caption{Results of dendrogram analysis.(Left) 3D dendrogram of G326 region extracted from $^{13}$CO(3--2) data showing hierarchical structures within the cloud. (Right) Overlay of $^{13}$CO intensity map with the leaf structures identified using the dendrogram analysis.
\label{fig_dendro}}
\end{figure*}

\begin{figure*}
\includegraphics[width=0.48 \textwidth]{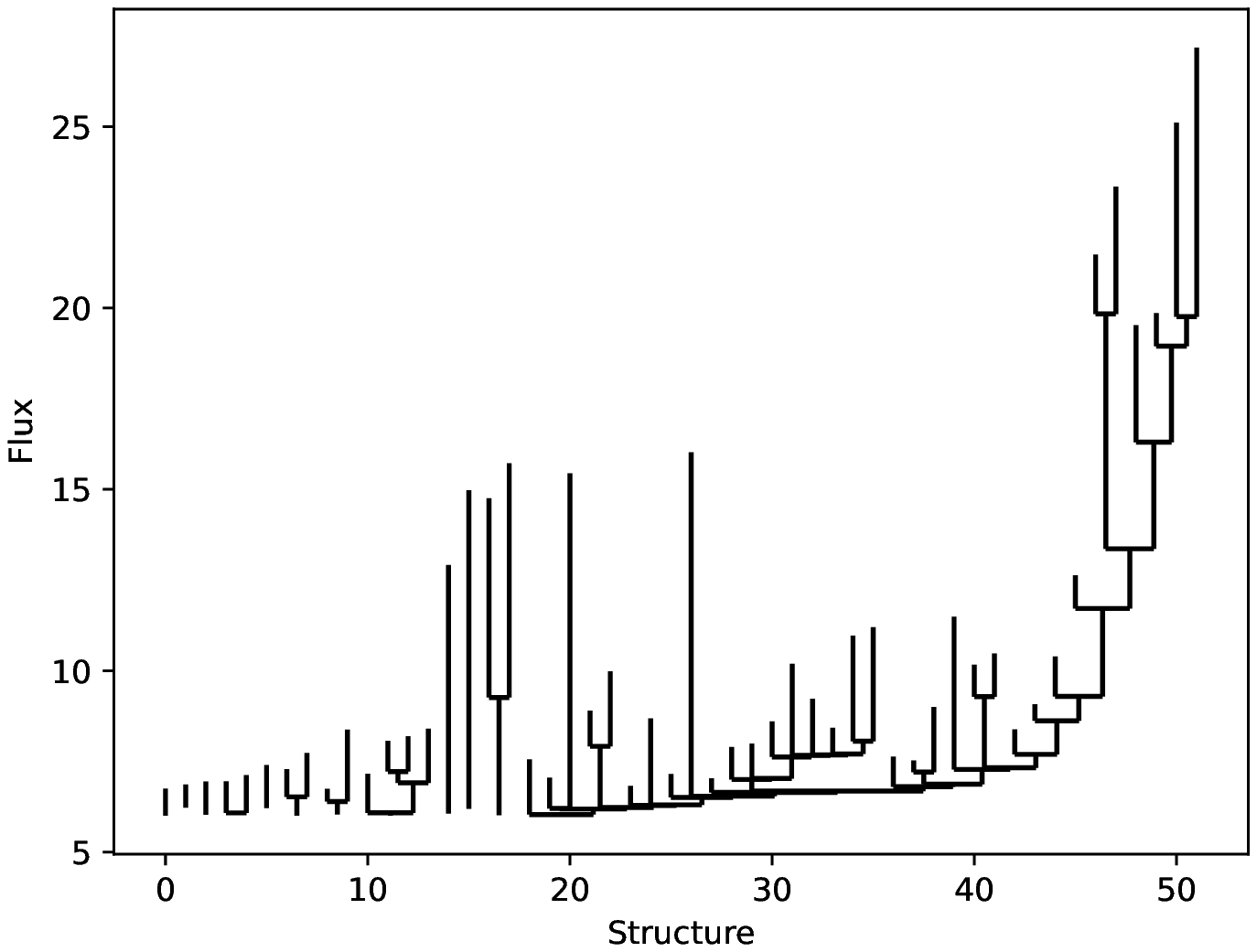}
\includegraphics[width=0.48\textwidth]{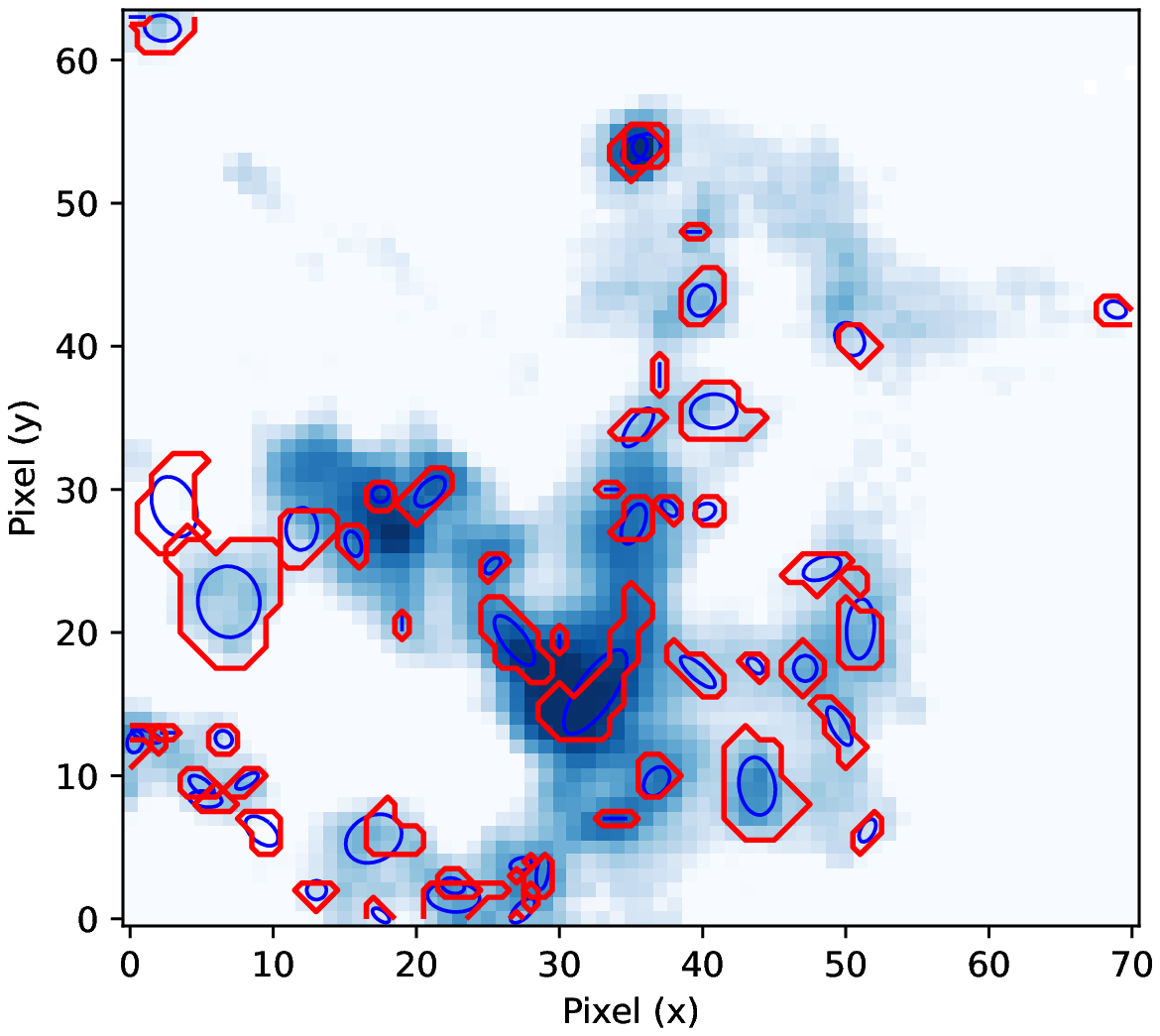}
\caption{Results of dendrogram analysis.(Left) 3D dendrogram of G326 region extracted from $^{12}$CO(3--2) data showing hierarchical structures within the cloud. (Right) Overlay of $^{12}$CO(3--2) intensity map with the leaf structures identified using the dendrogram analysis.
\label{fig_dendro2}}
\end{figure*}

The results of analysis of dendrogram analysis of the three-dimensional data cubes of \thCO(3--2) (Fig.\,\ref{fig_dendro}) and \twCO(3--2) (Fig.\,\ref{fig_dendro2}) to identify clumps in the G326 region are discussed in Sec.\,5.1. 

\begin{table*}
\caption{Properties of molecular clumps detected in \thCO(3--2) emission map extracted using \textit{astrodendro}. Association with the Hi-GAL sources in the region (Table\,\ref{tab_fir}) are also marked. The masses of all clumps, except for S6, S7 and S10 are derived from the integrated column density $N_{\rm cl}$(H$_2$) assuming $T_{\rm ex}=25$\,K and optical depth of \thCO(3--2) $\tau_{13}$=0.5 corresponding to a conversion factor of 4.4$\times 10^{20}$\,\cmsq /K\,\kms. The average density $n$(H$_2$) are calculated using the $M_{\rm cl}$ and radius.
\label{tab_dendro}}
\begin{tabular}{lrllrrrrrrrrrrrr}
\hline
\hline
CL\#& RA (J2000) & Dec(J2000) &  Radius & $I$(\thCO(3--2) & $\upsilon_{\rm LSR}$ & $\sigma_\upsilon$ & $N_{\rm cl}$(H$_2$) & $M_{\rm cl}$ &
$\alpha_{\rm vir}$ & Asso. & n(H$_2$)\\
 & & &   & & &  & $\times 10^{22}$ & &  &  &  $\times 10^4$\\
 & & &  (pc) & (K\,\kms) & (\kms) & (\kms) & (\cmsq) & (\msun) &  &  &\cmcub \\
\hline
1& 15:46:43.65 & -55:06:25.66 &  0.12  &   23.36  &   -64.42  &  0.86 & 1.0  & 7.6  & 14.6 &  & 2.1\\           
2& 15:46:46.77 & -55:06:00.86 &  0.10  &   16.14  &   -64.34  &  0.47 & 0.7  & 5.2  & 4.8 & S2 & 2.5\\         
3 & 15:46:47.76 & -55:09:36.98 &  0.08  &   12.27  &   -60.44  &  0.32 & 0.5  & 4.0   & 2.6 & &3.8\\           
4 & 15:46:48.13 & -55:06:52.63 &  0.19  &   37.90  &   -63.71  &  0.56 & 1.7  & 12.3 & 5.7 & & 0.9\\             
5 & 15:46:50.46 & -55:05:12.26 &  0.16  &   25.92  &   -71.42  &  0.53 & 1.1  & 8.4  & 6.2 & S3 & 1.0\\        
6 & 15:46:50.22 & -55:11:01.35 &  0.20  &   47.27  &   -57.81  &  1.04 & 2.1  & 15.3  & 17.1 & & 0.9\\         
7 & 15:46:54.52 & -55:12:08.83 &  0.16  &   55.73  &   -58.03  &  0.65 & 2.5  & 18.1  & 4.6 & & 2.1\\          
8 & 15:46:55.65 & -55:05:22.12 &  0.21  &   62.87  &   -72.19  &  0.61 & 2.8  & 20.4  & 4.5 & S4 & 1.1\\       
9 & 15:46:58.82 & -55:07:45.06 &  0.17  &   29.64  &   -61.75  &  0.50 & 1.3  & 9.6  & 5.2 & & 0.9\\           
10 & 15:47:00.07 & -55:10:47.76 &  0.23  &  112.32  &   -58.77  &  0.86 & 4.9  & 36.5  & 5.5 & & 1.5\\         
11 & 15:47:03.96 & -55:04:49.04 &  0.26  &  823.31  &   -72.92  &  1.74 & 36.2  & $^a$413.3 & 2.3 & S6 & 11.3\\ 
12 & 15:47:04.83 & -55:09:14.54 &  0.10  &   45.66  &   -62.24  &  0.50 & 2.0 & 14.8  & 2.0 & & 7.2\\           
13 & 15:47:05.50 & -55:12:29.95 &  0.13  &   30.97  &   -57.30  &  1.46 & 1.4 & 10.1  & 32.5 & & 2.2\\          
14 & 15:47:05.66 & -55:08:54.37 &  0.11  &   17.34  &   -58.72  &  0.63 & 0.8 & 5.6  & 9.4 & & 2.0\\            
15 & 15:47:10.23 & -55:11:14.36 &  0.24  &  589.19  &   -60.79  &  1.04 & 25.9  & $^b$187.2 & 1.6 & S7 & 6.6\\  
16 & 15:47:11.24 & -55:09:06.37 &  0.17  &   22.63  &   -71.93  &  0.50 & 1.0  & 7.4  & 6.9 & & 7.3\\           
17 & 15:47:12.98 & -55:12:24.31 &  0.14  &   29.27  &   -58.05  &  0.60 & 1.3  & 9.5  & 6.2 & & 1.7\\           
18 & 15:47:14.50 & -55:13:09.14 &  0.11  &   25.64  &   -61.08  &  0.82 & 1.1  & 8.3  & 10.5 & S9 & 3.0\\       
19 & 15:47:15.34 & -55:10:26.62 &  0.12  &  126.64  &   -60.35  &  0.46 & 5.6  & 41.1  & 0.7 & & 11.5\\         
20 & 15:47:15.35 & -55:13:43.39 &  0.19  &   29.17  &   -55.09  &  0.39 & 1.3  & 9.5  & 3.7 & & 6.7\\            
21 & 15:47:19.30 & -55:13:36.23 &  0.19  &   45.69  &   -56.29  &  0.49 & 2.0  & 14.8  & 3.7 & & 1.0\\           
22 & 15:47:25.27 & -55:09:05.44 &  0.46  & 1725.93  &   -60.55  &  1.62 & 75.9  & $^c$535.1 & 2.7 & S10 & 2.7\\  
23 & 15:47:26.39 & -55:12:46.06 &  0.17  &  100.08  &   -58.31  &  0.99 & 4.4  & 32.5  & 6.2 & & 3.2\\          
24 & 15:47:26.59 & -55:13:08.65 &  0.13  &   40.07  &   -56.33  &  0.58 & 1.8  & 13.0  & 4.0 & S12 & 2.9\\        
25 & 15:47:37.14 & -55:10:02.79 &  0.34  &  469.05  &   -62.18  &  1.14 & 20.6  & 153.4 & 3.5 & S15 & 1.9\\      
\hline
\hline
\end{tabular}

$^a$ $T_{\rm ex}$=15\,K, $\tau_{13}=0.7$, conversion factor of 6.8$\times 10^{20}$\,\cmsq /\Kkms.\\
$^b$ $T_{\rm ex}$=28\,K, $\tau_{13}=0.5$, conversion factor of 4.3$\times 10^{20}$\,\cmsq /\Kkms.\\
$^c$ $T_{\rm ex}$=24\,K, $\tau_{13}=0.4$, conversion factor of 4.2$\times 10^{20}$\,\cmsq /\Kkms.\\
\end{table*}

\section{Velocity Profile derived along the filaments \label{sec_velprof}}

\begin{figure*}
\includegraphics[width=0.48 \textwidth]{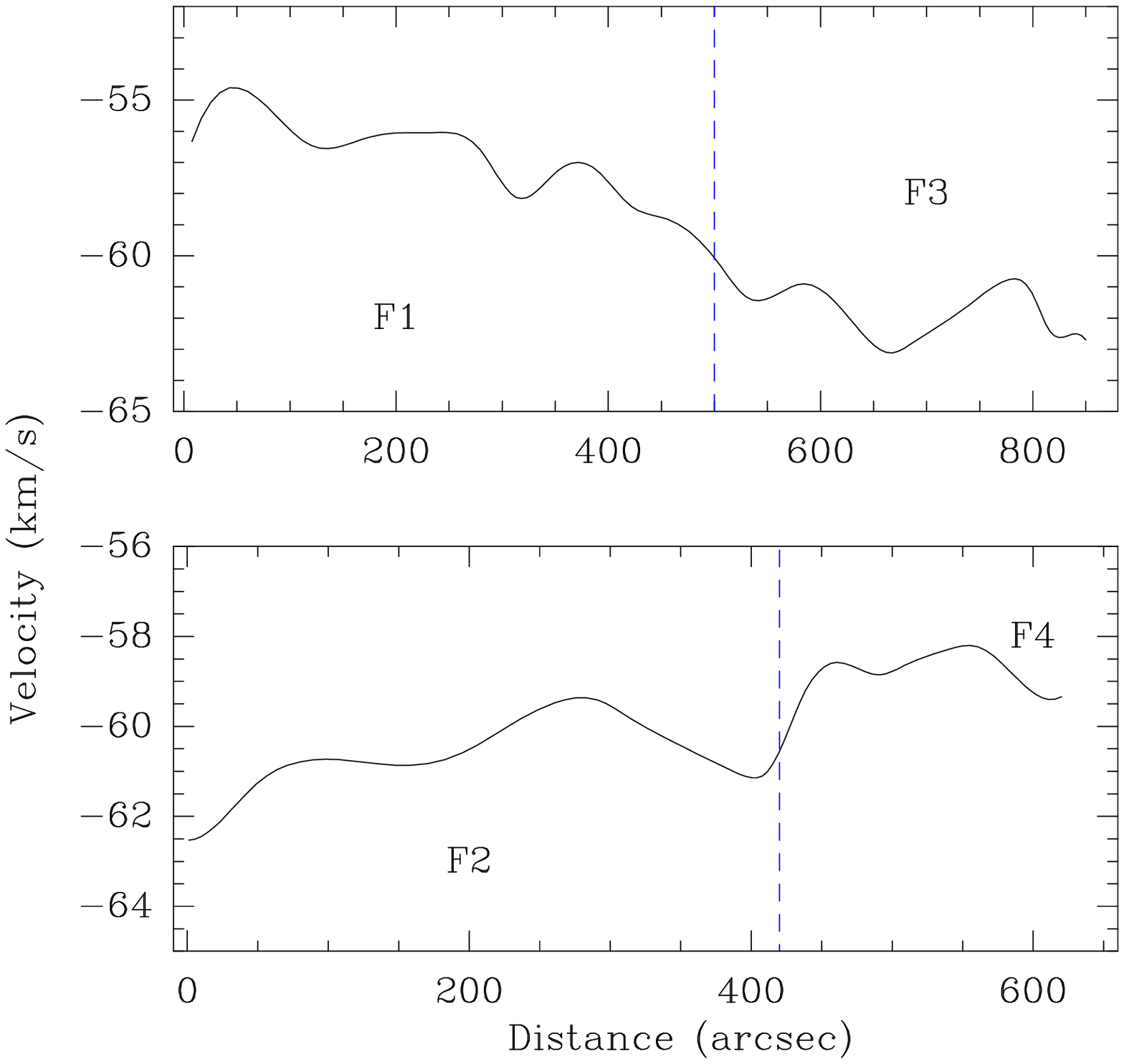}
\caption{Velocity profile derived at positions on the filaments F1, F2, F3 and F4 as marked in Fig.\,\ref{fig_pvdiag}. The accretion flows are estimated based on these profiles.
\label{fig_velprof}}
\end{figure*}

\label{lastpage}
\end{document}